\documentclass[letterpaper,10pt]{article}
\usepackage{osameet2}
\usepackage{graphicx}
\usepackage{subcaption}
\usepackage[font=small]{caption}
\usepackage{fancyvrb}

\usepackage{amsmath,amssymb}

\usepackage[pdftex,colorlinks=true,bookmarks=false,citecolor=blue,urlcolor=blue]{hyperref} 
\usepackage[utf8]{inputenc}

\usepackage{appendix}
\usepackage{soul}
 \definecolor{ngreen}{rgb}{0.2,0.6,0.2}

 \definecolor{npurple}{rgb}{0.8,0.2,0.8}

\begin{document}
 \clearpage\thispagestyle{empty}

\title{Orbital Angular Momentum (OAM) Mode Mixing in a Bent  Step Index  Fiber in Perturbation Theory}

\author{Ramesh Bhandari}
\address{Laboratory for Physical Sciences, 8050 Greenmead Drive, College Park,  Maryland 20740, USA}
\email{rbhandari@lps.umd.edu}

\begin{abstract}
Within the framework of perturbation theory, we explore in detail the mixing of orbital angular (OAM)  modes due to a fiber bend in a step-index multimode fiber. Using scalar wave equation, we  develop a complete set of analytic expressions for mode-mixing, including those for the $2\pi$ walk-off length, which is the distance traveled within the bent fiber before an OAM mode transforms  into its negative topological charge counterpart, and back into itself. The derived results provide insight into the nature of the bend effects, clearly revealing the mathematical dependence on the bend radius and the topological charge. We numerically simulate the theoretical results with applications to a few-mode fiber and a multimode fiber,   and calculate bend-induced  modal crosstalk with implications for  mode-multiplexed systems. The presented perturbation technique is general enough to be applicable to other perturbations like ellipticity  and  easily extendable to other fibers with step-index-like  profile as in the ring  fiber. 
\end{abstract}


\section{Introduction}
Ever since the revival of interest in orbital angular momentum (OAM) of light [1], research on OAM mode propagation  in a dielectric waveguide such as a multimode fiber has increased significantly. In commercial telecommunications, internet, and data centers, the orthogonality of the OAM modes leads to the possibility of multifold increase in traffic flow within a fiber by stacking traffic into the different OAM modes \cite{bozinovic, yue, huang,zhu,wang, rama,rama2}.   However,  a general drawback in practical fibers is the presence of imperfections such as ellipticity  and fiber bends, which mix these modes, and which then must be addressed in the analysis and design of fibers.

In this paper, we examine in detail the mixing of the OAM modes due to bends in a  step-index  fiber.  Previous studies of the impact of fiber bends have not explicitly considered OAM modes \cite{marcuse, garth, blake} and/or have been confined to a different type of fiber \cite{wang,rama,rama2}. Garth \cite{garth} provides a perturbative approach for the study of the modal fields in the presence of a bend; his work is primarily confined to the single mode fiber and  \emph{Linearly Polarized (LP)} modes corresponding to very low $l$ values. Chen and Wang \cite{wang} use a  finite-element vector wave equation solver to study the fiber bend effect for a graded-index fiber.  Gregg et al. \cite{rama, rama2} employ small bend angles \cite{blake} to obtain a  general trend in the difficulty of  mixing of the degenerate modes  due to bends in an air-core fiber. In this work, we provide a complete formalism for the mixing of OAM modes and crosstalk in a step-index fiber using perturbation theory. The use of perturbation theory provides insight into the mechanics of mixing, leading to  selection rules and analytic expressions, including those for degenerate mode-mixing; the extent of degenerate mode mixing is defined through a $2\pi$ walk-off length parameter, frequently cited in literature  \cite{wang,yue}.

In what follows, we invoke the weakly-guiding approximation (WGA) due to the fact that the  refractive index of the core is only slightly greater than the cladding refractive index in step-index fibers, and employ the scalar wave equation in the solution of our problem. 

\section{OAM Modes in a Straight Fiber}
We denote the refractive index of the core and the cladding by $n_1$ and $n_2$ respectively.  In WGA, where $(n_1^2-n_2^2)/(2n_1)<<1$,  the vector mode solutions $HE_{l+1,m}, HE_{-l-1,m},EH_{l-1,m},$ and $EH_{-l+1,m}$ ($l>1$ ),  when expressed in the $e^{\pm il \theta}$ azimuthal basis (see, e.g., \cite{yariv}), collapse respectively into a single set of four degenerate (scalar) modes:  $\phi_{l,m}\vec{\epsilon}_+, \phi_{-l,m}\vec{\epsilon}_-, \phi_{l,m}\vec{\epsilon_-}, \phi_{-l,m}\vec{\epsilon_+}$
(for $l=1$, which is a special case, $HE_{2,m}, HE_{-2,m}, 1/\sqrt{2}(TM_{0,m}+iTE_{0,m}), 1/\sqrt{2}(TM_{0,m}-iTE_{0,m})$ reduce respectively to $\phi_{1,m}\vec{\epsilon}_+, \phi_{-1,m}\vec{\epsilon}_-, \phi_{1,m}\vec{\epsilon_-}, \phi_{-1,m}\vec{\epsilon_+}$; see also \cite{rama3}).  Each mode within the scalar quartet has the same propagation constant, denoted by $\beta_{l,m}$.  
{ $\vec{\epsilon}_\pm = 1/\sqrt{2}(\hat{x}\pm i\hat{y})$ represents left-circularly (+ sign) polarized light and right-circularly (- sign) polarized light,  which correspond respectively to spin $S=1$ and spin $S=-1$.  Spin $S$ has the physical meaning of each photon within the mode carrying a  spin angular momentum of $S\hbar$.  Similarly, $l$ denotes the topological charge, and implies an OAM value of $l\hbar$ per photon within that mode. Within the vector quartet above, the spin S  manifests itself in the four indices, $\pm l\pm 1$. They are the result of the angular momentum addition, $J=l+S$, where $J\hbar$ may be regarded as the total angular momentum per photon within that mode.   
The spatial wave function, $\phi_{l,m}$, is written as 
\begin{equation}
 \phi_{l,m}   (r,\theta,z) =O_{l,m}(r,\theta) e^{i\beta_{l,m} z};
\end{equation}
 $r,\theta$, and $z$ are cylindrical polar coordinates with the z axis coincident with the fiber axis  (see Fig. 1a); the wave is propagating in the +z direction (out of the plane of the paper).  $O_{l,m}$, the amplitude, is an eigenvalue solution of the scalar wave equation \cite{snyder}:
\begin{equation}
HO_{l,m}(r,\theta)=\beta_{l,m}^2 O_{l,m}(r,\theta).
\end{equation}
\begin{footnotesize}
\begin{figure}[htbp]
  \centering
  \includegraphics[width=11cm]{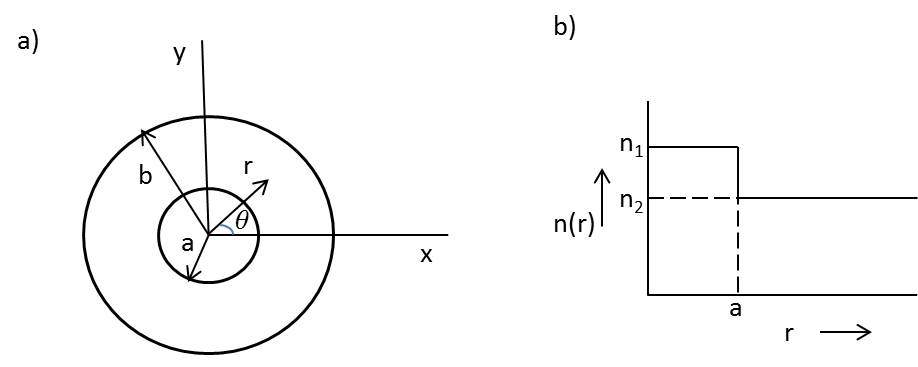}
\caption {  a) Cross section of the step-index fiber with core radius $a$ and cladding radius $b (>>a)$; $r$ and $\theta$ along with the $z$ coordinate (z axis coincident with the fiber axis) constitute the cylindrical polar coordinates; the  mode is assumed propagating in the +z direction (out of the plane of the paper). b) The step-index profile, where $n_1$ and $n_2$ denote the refractive index of the core and cladding, respectively.}
\end{figure}
\end{footnotesize}
 The Hermitian operator $H$ is equal to $\vec{\nabla}_t^2+k^2n_1^2$ for $r\le a$ (core radius) and equal to $\vec{\nabla}_t^2+k^2n_2^2$ for $r\ge a$   (see Fig. 1b);  $k=2\pi/\lambda$, where $\lambda$ is the wavelength; $\vec{\nabla}_t^2$ is the  transverse Laplacian:   $\partial^2/\partial r^2 + (1/r) \partial/\partial r +(1/r^2) \partial^2/\partial \theta^2 $.
The amplitude, $O_{l,m}(r,\theta)$, of the electric field is given by  
\begin{align}
O_{l,m}(r,\theta)=\frac{1}{\sqrt{N_{l,m}}}J_l(p_{l,m}r)e^{il\theta} \quad for\ r\le a \nonumber \\
=\frac{1}{\sqrt{N_{l,m}}}\frac{J_l(p_{l,m}a)}{K_l(q_{l,m}a)}K_l(q_{l,m}r)e^{il\theta} \quad for \ r\ge a.
\end{align}
$N_{l,m}$ is the normalization constant which can be determined analytically from the properties of the Bessel functions \cite{mw}; $p_{l,m}=\sqrt{k^2 n_1^2-\beta_{l,m}^2}$ and $q_{l,m}=\sqrt{\beta_{l,m}^2-k^2 n_2^2}$.   $O_{l,m}(r,\theta)$ characterized by an exponential azimuthal dependence, is referred to as the amplitude   (or the field profile)  of an OAM mode corresponding to a topological charge $l$ and a radial mode number $m$; hereafter we denote such a mode by $OAM_{l,m}$ .  The wave solution is continuous at the boundary $r=a$; the further requirement of the continuity of the first (radial) derivative at $r=a$ then gives the characteristic equation from which the wave propagation constants $\beta_{l,m}$ are computed  \cite{snyder}. The amplitude $O_{-l,m}(r,\theta)$ of the degenerate $-l$ solution ($\beta_{-l,m}=\beta_{l,m}$), is also given by Eq. 3, except that $e^{il\theta}$ is replaced with $e^{-il\theta}$.
\\\\
Within WGA, as discussed above, the vector modes reduce to scalar modes, which are the products of the spatial mode, the $OAM_{l,m}$ defined above and the polarization, $\epsilon_\pm$. The impact of a bend is then the product of the  impact on the spatial mode and the impact on the polarization  (birefringence), evaluated separately.  The latter has been studied in detail in connection with single-mode fibers ($l=0$) in the past (see, e.g., \cite{smith}). The primary purpose of this paper is to provide a detailed, quantitative treatment of the impact of the bend on the spatial $OAM$ modes, as manifested in their mixing and the crosstalk they generate. We, therefore, do not concern ourselves here with the explicit calculation of the polarization changes induced by the bend. In the next section, focusing on the spatial  modes, we formulate the scalar wave equation in the presence of a bend.  
\section{ Scalar Wave Equation for the Bent Fiber}
When a straight fiber is bent into a fiber of radius $R>>a$, wave propagation through the bent fiber is modeled as  wave propagation through a straight fiber with an equivalent refractive index profile given by \cite{marcuse, marcuse2}
\begin{equation}
n_e^2(r,\theta)=n'^2 +2 n'^2(r/R)\cos\theta
\end{equation}
where $n'^2=n_1^2$ for $r\le a$ and $n'^2=n_2^2$ for $r\ge a$ (see Figure 1b).   The equivalent refractive index depends upon the radial distance $r$ and the azimuthal angle $\theta$. Considering the bend in Figure 2, and referring to Figure 1a, $\theta=0$  corresponds to the outer edge of  the straightened bend, where the equivalent refractive index is the largest (see Eq. 4), while $\theta =\pi$ in Figure 1a corresponds to the inner edge, where the equivalent refractive index is the smallest.
%
%
%
%
%
%
\\\\
Modifying the $H$ operator in Eq. 2 to include the bend-induced correction term (the second term of Eq. 4), we obtain a perturbed wave equation for the straight fiber:
\begin{equation}
(H+\lambda'\delta H)O_{l,m}'(r,\theta)=\beta_{l,m}'^2O_{l,m}'(r,\theta),
\end{equation}
where the perturbation parameter
$\lambda'=a/R<< 1$
and $\delta H=(2k^2n'^2r\cos\theta)/a$; $\beta'_{l,m}$ and $O'_{l,m}$ are respectively the perturbed propagation constant and OAM amplitude. The $O'_{l,m}$'s, like the $O_{l,m}$'s, form a complete orthonormal set;   they are the  eigenfunctions of the perturbed Hermitian operator, $H+\lambda'\delta H$.  In what follows, we assume that the perturbation is small ($\lambda' \delta H<<H$) and use perturbation theory to solve Eq. 5; we also suppress the arguments $r,\theta$ for convenience, unless required by the context.

\section{Perturbation Solution for the Bent Fiber}
 We expand the perturbed amplitude $O_{l,m}'$ and the perturbed propagation constant $\beta_{l,m}'$,   using the standard techniques of perturbation theory \cite{landau,mw,soliverez},  as
\begin{equation}
O_{l,m}'=O_{l,m}+\sum'_{n,k} a_{(l,m)(n,k)}^{(1)}O_{n,k}+\sum'_{n,k} a_{(l,m)(n,k)}^{(2)}O_{n,k}+....,
\end{equation}
\begin{equation}
\beta_{l,m}'^2=\beta_{l,m}^2+\beta_{l,m}^{2(1)}+\beta_{l,m}^{2(2)}+....,
\end{equation}
where the contributions in different orders of the perturbation parameter $\lambda'$ are indicated by the superscripts in parentheses, and the prime on the summation implies  $n\ne l$.
Subsequently, we insert the above series in Eq. 5, take the necessary scalar products to solve in different perturbation orders \cite{landau,mw,soliverez}, and obtain the analytic expressions   for the mixing coefficients and the propagation constant corrections:
\begin{equation}
 a_{(l,m)(n,k)}^{(1)}=\lambda'\frac{\delta H_{(n,k)(l,m)}}{(\beta_{l,m}^2-\beta_{n,k}^2)}, 
\end{equation}
\begin{equation}
 a_{(l,m)(n,k)}^{(2)}=\lambda'^2\bigg( \sum_{r,s}\frac{\delta H_{(n,k)(r,s)}\delta H_{(r,s)(l,m)}}{(\beta_{l,m}^2-\beta_{n,k}^2)(\beta_{l,m}^2-\beta_{r,s}^2)}-\frac{\delta H_{(n,k)(l,m)}\delta H_{(l,m)(l,m)}}{(\beta_{l,m}^2-\beta_{n,k}^2)^2}\bigg), 
\end{equation}
%
\begin{equation}
\beta^{2(1)}_{l,m}=\lambda'\delta H_{(l,m)(l,m)},
\end{equation}
\begin{equation}
\beta^{2(2)}_{l,m}=\lambda^{'2}\sum_{n,k}\frac{\delta H_{(l,m)(n,k)}\delta H_{(n,k)(l,m)}}{(\beta_{l,m}^2-\beta_{n,k}^2)},
\end{equation}
%
up to second order (higher order contributions, although more complicated, can similarly be determined, if needed).
The matrix  element, $\delta H_{(l',m')(l,m)}$ is a scalar (inner) product defined as
\begin{equation}
\delta H_{(l',m')(l,m)}=<O_{l',m'}|\delta H|O_{l,m}>=\int O_{l',m'}^* (\delta H) O_{l,m} rdr d\theta= \frac{2k^2 n'^2}{a}\int\int O_{l',m'}^* (r\cos\theta) O_{l,m} rdr d\theta;
\end{equation}
the bra ($<$), ket ($>$) notation signifies a scalar product.   This matrix element times $\lambda'$ represents the bend-induced interaction (or coupling) between the $OAM_{l,m}$ and $OAM_{l',m'}$ modes.

\subsection{Selection   Rule and the   Simplification of the   Analytic   Expressions}
 Recalling the azimuthal dependence of $O_{l,m}$ (see Eq. 3), we immediately note that the  perturbation matrix element, $\delta H_{(l',m')(l,m)}$ given in the above equation
is non-zero only when $|\Delta l|=|l'-l|=1$, i.e.,
\begin{equation}
\delta H_{(l',m')(l,m)}= \delta H_{(l',m')(l,m)}\delta_{l',l+1}+\delta H_{(l',m')(l,m)}\delta_{l',l-1},
\end{equation}
where the second $\delta$ in each of the above two terms on the right-hand side is the Kronecker delta: $\delta_{i,j}=1$  if $i=j$, otherwise 0.
This selection rule then implies that an OAM  mode's amplitude perturbed by a fiber bend acquires (in first order perturbation)  an admixture of other modes' amplitudes only if its topological charge $l$ differs from that of the admixed modes by $\pm 1$.  Substitution of Eq.13  simplifies  the analytic expressions and  a pattern emerges from which mixing   of amplitudes of different topological charges, to  lowest order  in perturbation, can be written out:
\begin{equation}
 a_{(l,m)(l\pm1),n}^{(1)}=\lambda'\frac{\delta H_{(l\pm1,n)(l,m)}}{(\beta_{l,m}^2-\beta_{l\pm1,n}^2)}, 
\end{equation}
\begin{equation}
 a_{(l,m)(l\pm2,n)}^{(2)}=\lambda'^2 \sum_k\frac{\delta H_{(l\pm2,n)(l\pm1,k)}\delta H_{(l\pm1,k)(l,m)}}{(\beta_{l,m}^2-\beta_{l\pm2,n}^2)(\beta_{l,m}^2-\beta_{l\pm1,k}^2)}, 
\end{equation}
\begin{equation}
 a_{(l,m)(l\pm3,n)}^{(3)}=\lambda'^3 \sum_{j,i}\frac{\delta H_{(l\pm3,n)(l\pm2,j)}\delta H_{(l\pm2,j)(l\pm 1,i)}\delta H_{(l\pm1,i)(l,m)}}{(\beta_{l,m}^2-\beta_{l\pm3,n}^2)(\beta_{l,m}^2-\beta_{l\pm2,j}^2)(\beta_{l,m}^2-\beta_{l\pm1,i}^2)},
\end{equation}
and so on.  The expressions for the propagation constants in Eqs. 10 and 11  become
\begin{equation}
\beta^{2(1)}_{l,m}=0,
\end{equation}
\begin{equation}
\beta^{2(2)}_{l,m}=\lambda'^2\bigg(\sum_k\frac{\delta H_{(l,m)(l+1,k)}\delta H_{(l+1,k)(l,m)}}{(\beta_{l,m}^2-\beta_{l+1,k}^2)}+\sum_{k'}\frac{\delta H_{(l,m)(l-1,k')}\delta H_{(l-1,k')(l,m)}}{(\beta_{l,m}^2-\beta_{l-1,k'}^2)}\bigg).
\end{equation}

As an example, in first order perturbation (Eq. 14), an input $OAM_{3,m}$ mode will couple with  $l=2$ or $l=4$ OAM modes with radial index $n\ge1$.   Not discernible in the above equations, the selection rule, $\Delta l=\pm 1$, however, also allows the following scenario: $OAM_{3,m}$ mode couples first to an $l=4$ mode, which in turn couples back to $OAM_{3,m}$ mode, which subsequently couples to the $l=2 $ mode. This  coupling to the $l=2$ OAM mode takes place in three steps, or in third order perturbation.  Since each step involves a factor of $\lambda'$, the contribution of this term to the mixing coefficient of the admixed $l=2$ mode  is of $\lambda'^3$, as compared to the contribution of order $\lambda'$ originating in the direct single step (or lowest order) coupling described by Eq. 14. Since $\lambda'<<1$ (see Section 3), such higher order contributions to the mixing coefficient of an admixed  mode of a given topological charge are smaller by at least a power of $\lambda'^2$, and thus  negligibly small compared to the direct minimal number of $\Delta l=\pm 1$ step contributions exhibited in Eqs. 14-16. They are, therefore, not included within the above expressions, Eqs. 14-16. Subsequently, in accordance with Eq. 15, 
the incoming $OAM_{3,m}$ mode will also couple to $l=1$ mode (any allowed $n$ value), however, in two steps (second order perturbation) via the intermediate  $l=2$ modes (any allowed radial index $k$); it will similarly couple to $l=5$ modes   in second order perturbation via  the intermediate $l=4$  modes. The third order perturbation result, Eq. 16, will couple the incoming $l=3$ mode to $l=0$ or $l=6$ modes in three steps of cumulative strength, $(\lambda')^3$ via a number of intermediate modes characterized by radial indices, $i$ and $j$.
 In general, two OAM modes differing in  their topological charge values by $q$ can only mix in $|q|_{th}$ order perturbation,   i.e., the mixing coefficient is of order $(\lambda')^{|q|}$ in strength. Thus, greater the value of $|q|$, weaker the mixing tends to be. From Eqs. 14-16, we also see that larger the propagation constant differences, smaller the values of the mixing coefficients.
%
%
\subsection{Results for Negative Topological Charge, $-l$}
In what follows, we indicate modes with negative topological charge by placing an explicit minus sign in front of $l$, and assume $l$ is always positive. An $OAM$ mode with topological charge $-l$ and mode number $m$ is then denoted by $OAM_{-l,m}$.  The perturbed counterpart of the $OAM_{-l,m}$ mode is described by the same series expansion as in Eq. 6, except that $-l$ replaces $l$ everywhere:
\begin{equation}
O_{-l,m}'=O_{-l,m}+\sum'_{n,k} a_{(-l,m)(n,k)}^{(1)}O_{n,k}+\sum'_{n,k} a_{(-l,m)(n,k)}^{(2)}O_{n,k}+.....
\end{equation}
 The  mixing coefficients obtained by flipping the signs of topological charge indices in Eqs.14-16 are identical in values to their counterparts  given in  Eqs. 14-16 for the $+l$ case. This is due to  the fact that $\beta_{-l,m}=\beta_{l,m}$ (the degeneracy between the $OAM_{l,m}$ and the $OAM_{-l,m}$ modes) as well as the property
\begin{equation}
 \delta H_{(-l',m')(-l,m)}=\delta H_{(l',m')(l,m)},
\end{equation}
which follows from Eqs. 12 and 3. This, however, is not surprising, given the symmetry of the wave equation with respect to $l$ and $-l$.
 Note also from Eqs. 12 and 3 that the $\delta H$ matrix elements are real and symmetric $ \delta H_{(l,m)(l',m')}=\delta H_{(l',m')(l,m)}$ , consistent with the Hermitian property of $\delta H$. 
\subsection{Mixing of Degenerate Mode Amplitudes, $O_{l,m}$ and $O_{-l,m}$}
We must now address the question of degeneracy between $OAM_{l,m}$ and $OAM_{-l,m}$ modes ($ l>0$) . From arguments of symmetry, we expect the two  diagonal elements of the $2x2$  matrix pertaining to the $(l,-l)$ subspace to be equal to each other; similarly,  the off-diagonal elements connecting the $l$ state to the $-l$ state and vice versa should be equal to each other, and not necessarily the same as the diagonal elements; furthermore, from the selection rule, $\Delta l =\pm 1$ , we immediately see that the $OAM_{l,m}$ and $OAM_{-l,m}$ modes can only be connected in $2l$ steps: $l\rightarrow l-1\rightarrow l-2...1\rightarrow 0\rightarrow -1 ...-(l-2)\rightarrow -(l-1)\rightarrow \-l$, i.e., the off-diagonal elements become nonzero in perturbation order $2l$. In Appendix A, we describe and carry out the formal procedure  to obtain analytic expressions for the breaking of the degeneracy. The new eigenvalues are of the form:
\begin{equation}
(\beta'^{\pm}_{l,m})^2=\beta^2_{l,m}+\epsilon_{l,m} \pm \Delta\beta'^2_{l,m}/2;
\end{equation}
$ \epsilon_{l,m}= \gamma^{(2l-1)}, \Delta\beta'^2_{l,m}=2\kappa^{(2l-1)}$, where these parameters are evaluated in Appendix A.  The eigenvalue difference, $\Delta\beta'^2_{l,m}$, is given explicitly  by
\begin{equation}
\Delta\beta'^2_{l,m}=
2(\lambda')^{2l}\sum\frac{\delta H_{(l,m)(l-1,n)}\delta H_{(l-1,n)( l-2,k)}......\delta H_{(1,i)(0,j)}\delta H_{(0,j)(-1,k)}....\delta H_{(-l+2,r)(-l+1,s)}\delta H_{(-l+1,s)(-l,m)}}{(\beta_{l,m}^2-\beta_{l-1,n}^2)(\beta_{l,m}^2-\beta_{l-2,k}^2)........(\beta_{l,m}^2-\beta_{0,j}^2)...........(\beta_{l,m}^2-\beta_{-l+2,r}^2)(\beta_{l,m}^2-\beta_{-l+1,s}^2)}. 
\end{equation}
The right-hand sum in Eq. 22 runs over all radial order solutions of the intermediate coupled modes.
The corresponding eigenamplitudes are (Appendix A)
\begin{equation}
O^\pm_{l,m}=1/\sqrt{2}(O_{l,m}\pm O_{-l,m}).
\end{equation} 
These linear combinations can be interpreted as the  field amplitudes of the corresponding \emph{Linearly Polarized (LP)} modes \cite{buck},  denoted $LP^{(e)}_{l,m}$ (for the $+$ sign in Eq. 23) and  $LP^{(o)}_{l,m}$ for the - sign, with intensity proportional to $\cos^2l\theta$ and $\sin^2l\theta$, respectively (generally the notation $a,b$ is used in the literature for the degenerate $LP_{l,m}$ pair, but we use the notation $e$ (for even) and $o$ (for odd) to avoid a conflict with the notation $b$ introduced below for the bend). Note, however, that the $O_{l,m}^-$ combination or the corresponding $LP^{(o)}_{l,m}$ mode amplitude has an extra factor of complex $i$ (see Eq. 3); alternatively, it carries an extra phase of $\pi/2$ with respect to the $O^+_{l,m}$ amplitude.  Normally, in a straight fiber, these linear combinations would be degenerate (propagation constant equal to $\beta_{l,m}$), but the bend effect causes them to become slightly nondegenerate with respect to each other (see Eqs. 21 and 22). Inverting Eq. 23, we find 
\begin{equation}
O_{\pm l,m}=1/\sqrt{2}(O^+_{l,m}\pm O^-_{l,m}).
\end{equation}
What this physically means is that an $OAM_{l,m}$ mode, $\phi_{l,m}(z)=O_{l,m}e^{i\beta_{l,m}}(z)$ (see Eq. 1), entering a bend of radius $R$ at $z=0$   travels down the fiber as an equal superposition of the two slightly  nondegenerate propagating $LP$ modal fields with propagation constants, $\beta'^+_{l,m}$ and $\beta'^-_{l,m}$,  determined from Eqs. 21 and 22.   Upon traversal of length $L$ within the bend,   its spatial wave function  becomes (\emph{ignoring interactions with other modes}) 
\begin{equation}
 \phi^{(b)}_{l,m}(L)= \frac{1}{\sqrt{2}}\Big(O_{l,m}^+e^{i\beta'^+_{l,m}L}+O^-_{l,m}e^{i\beta'^-_{l,m}L}\Big),
\end{equation}
where superscript $b$ (in parentheses) signifies the bend effect.   Substituting Eq. 23 in Eq. 25 and  using the fact that $\beta'^+_{l,m}+\beta'^-_{l,m}\approx 2\beta_{l,m}$ due to $\epsilon_{l,m}<<\beta_{l,m}$ (see Eq. 21), Eq. 25 simplifies to
\begin{equation}
\phi^{(b)}_{l,m}(L)=\Big(\cos((\beta'^+_{l,m}-\beta'^-_{l,m})L/2)O_{l,m}+ i \sin((\beta'^+_{l,m}-\beta'^-_{l,m})L/2) O_{-l,m}\Big)e^{i\beta_{l,m}L}.
\end{equation}
Thus, the incoming $OAM_{l,m}$ mode becomes a mixture of $OAM_{l,m}$ and $OAM_{-l,m}$ modes, with the modulus square of the mixing coefficients (cosine and sine factors) adding to unity; at $L=0$, $\phi^{(b)}_{l,m}(L)$ correctly reduces to $\phi_{l,m}(0)=O_{l,m}$, the amplitude of the input mode, $OAM_{l,m}$ . We remark here that the bend will also change the  polarization of the  input $OAM_{l,m}$ mode. For example, an initial left (or right) circular polarization will become a linear combination of left and right circular polarizations, which when multiplied with the right-hand-side of Eq. 26, will yield the final state in the combined spatial-spin domain (this is briefly discussed further in Section 7.1). However, as mentioned in Section 2, a detailed study of the bend effect on polarization is not within the scope of the current work; we deal here with the mixing of the spatial $OAM$ modes only.
\\\\
From Eq. 26, we see that the entering $OAM_{l,m}$ mode oscillates into and out of the $OAM_{-l,m}$ mode with a $2\pi$ walk-off length given by
\begin{equation}
L^{(2\pi)}_{l,m}=\frac{2\pi}{|\beta'^+_{l,m}-\beta'^-_{l,m}|};
\end{equation}
the $2\pi$ walk-off length is the distance over which a given  mode transforms into its degenerate partner, and back into itself. 
Substituting
\begin{equation}
 \beta'^\pm_{l,m}\approx \beta_{l,m}+\epsilon_{l,m}/(2\beta_{l,m})\pm \Delta\beta'^2_{l,m}/(4\beta_{l,m}),
\end{equation}
which follows from Eq. 21, we obtain explicit analytic expressions for the $2\pi$ walk-off length:
\begin{equation}
L^{(2\pi)}_{l,m}=\frac{4\pi\beta_{l,m}}{|\Delta\beta'^2_{l,m}|},
\end{equation}
where $\Delta\beta'^{2}_{l,m}$ is given by Eq. 22. The modulus square of the amplitudes within the output behaves as $\cos^2(\pi L/L^{(2\pi)}_{l,m})$ for the $OAM_{l,m}$ mode and $\sin^2(\pi L/L^{(2\pi)}_{l,m})$ for the $OAM_{-l,m}$ mode, with the former transitioning into its $-l$ counterpart at
 $L=L^{(2\pi)}_{l,m}(n+1/2)$, where $n\ge0$ is an integer. However, at  $L=L^{(2\pi)}_{l,m}(n+1/4)$, integer $ n\ge 0$, we find from Eqs. 26 and 27 that the amplitude of the admixed $OAM_{-l,m}$ mode has the same magnitude as the magnitude of the  input $OAM_{l,m}$ mode. Consequently, their interference produces an intensity at the output, $|\phi^{(b)}_{l,m}(L)|^2 \propto \cos^2(l\theta\pm\pi/4)$, depending upon the sign of $\beta'^+_{l,m}-\beta'^-_{l,m}$ in Eq. 26 (in the radial direction the intensity varies as the square of the Bessel functions  in Eq. 3). Interestingly, this result is nothing but the familiar $2l$-lobe intensity pattern of the $LP_{l,m}$ modes, rotated, however, by an angle $\pm\pi/(4l)$. This means for an input $OAM_{l,1}$ mode, depending upon the length $L$, we can expect to see an intensity pattern at the output, which varies from being  donut shaped (corresponding to the  pure OAM modes, with topological charge $l$ or $-l$)  to one with a $2l$-lobe pattern of the corresponding \emph{LP} mode, rotated by angle $\pm\pi/(4l)$. For example, for $l=1$, this angle of rotation is $\pi/4$ and for $l=2$ the angle of rotation is $\pi/8$. For an input $OAM_{-l,m}$ mode, we replace $l$ with $-l$ in the cosine-squared term above; the tilt is then in the opposite direction.  
\\\\
From Eqs. 29 and 22, we further see that for fixed $l$, the $2\pi$ walk-off length varies as $R^{2l}$. Thus, as $R$ increases, so does the $2\pi$ walk-off length, which approaches infinity as the bend flattens to become straight (case of $R=\infty$). Likewise, for fixed $R$, the $2\pi$ walk-off length grows exponentially as a function of $l$ due to the dominant multiplying factor, $\lambda'^{2l}$  in Eq. 22. This rapid rise in the value of the $2\pi$ walk-off length with $l$ is a reflection of the increasing difficulty in the ability of the the given bend (of radius $R$) to engender an OAM transfer of $2l$ as $l$ is increased (a transformation from an $OAM_{l,m}$ mode to an $OAM_{-l,m}$ mode implies a change of $2l$ in the value of OAM). 
\subsection{Perturbation Series Revisited} 

The appropriate linear combinations of the degenerate mode amplitudes, $O_{\pm l,m}$ (which are fixed by the fiber bend; see  Appendix A and Section 4.3) are  given by Eq. 23. We must work  with these linear combinations (the modal fields of the corresponding degenerate $LP_{l,m}$ pair) and replace the perturbation series for $O'_{\pm l,m}$ (Eqs. 6 and 19) with
\begin{equation}
O'^{\pm}_{l,m}=O^{\pm}_{l,m}+\sum'_{n,k} a^{(1)}_{(l,m)(n,k)}O^{\pm}_{n,k}+\sum'_{n,k} a^{(2)}_{(l,m)(n,k)}O^{\pm}_{n,k}+....,
\end{equation}
which we arrived at in Appendix A; the prime on the summations indicates $n \ne \pm l$.    Eq. 30 is now a perturbation series expressed completely in terms of the eigenamplitudes of the bent fiber. $O^{\pm}_{n,k}$, like the $ O^{\pm}_{l,m}$ amplitudes, are written as $1/\sqrt{2}(O_{n,k}\pm O_{-n,k})$.  Physically, Eq. 30 is then an expression of the bend-perturbed amplitude profile of an input $LP_{l,m}$ mode in terms of the unperturbed \emph{LP} mode amplitudes. The mixing coefficients for these $LP_{n,k}$ modes, $n \ne l$, are the same (equal to $a^{(i)}_{(l,m)(n,k)}$) as in the original perturbation series, Eqs. 6 and 19, for the $O'_{\pm l,m}$ amplitudes. This is not surprising, since the $ LP_{l,m}$ modal amplitudes are linearly related to the corresponding $O_{\pm l,m}$ amplitudes (Eq. 23), and the scalar wave equation is linear in the $OAM$ modal amplitudes.  Furthermore, because the derivation in Appendix A is valid for arbitrary values of topological charge, the degenerate combinations,  $O^{\pm }_{n,k}$, are also nondegenerate on account of the bend.   The propagation constants of the two slightly nondegenerate modes with amplitudes, $O^{\pm}_{n,k}$, are given by the same expressions as those in Eqs. A.19-23, except that $n$ replaces $l$, and $k$ replaces $m$; the dummy index $k$  in Eqs. A. 21-23 can be replaced with a different dummy index, say $k'$ in order to avoid confusion.  
 Note that $O^{'+}_{l,m}$ amplitudes expand in terms of the $O^{+}_{l',m'}$ amplitudes,  and similarly the  $O^{'-}_{l,m}$ amplitudes in terms of their corresponding counterparts, $O^{-}_{l',m'}$. Conversely, the $O^+_{l,m}$ amplitudes are expandable in terms of $O^{'+}_{l'm'}$ amplitudes  and similarly $O^-_{l,m}$ in terms of $O^{'-}_{l',m'}$;   this is consistent with the fact that the latter, which are the solutions of the perturbed wave equation, Eq. 5, form a complete set.   Hereafter,  the eigenmodes  corresponding to amplitudes, $O^{\pm}_{l',m'}$ will  be denoted by $OAM^{\pm}_{l',m'}$.
\\\\
In the next section, using the above  perturbation series, we now determine the complete expressions that include the mixing of the given input $OAM_{l,m}$ mode with the other modes $l'\ne -l$, not considered in Section 4.3. 

\section{OAM Mode-Mixing}
  For purposes of derivation,  consider now  an input $ OAM^+_{l,m}$  mode  traveling in a straight fiber before encountering a bend of radius $R$, which it then exits after traveling a length $L$ (see Figure 2).   Its  polarization may be assumed circular as before, although it does not matter since the scalar wave equation is independent of polarization, and we are working in the spatial domain only. The field amplitude $O^+_{l,m}$ of this spatial mode  corresponds to the \emph{LP}$^{(e)}_{l,m}$ mode (see discussion following Eq. 23). Let $\phi_{l,m}^{+(b)}(z)$ denote the spatial wave function at the coordinate $z$ within the bent fiber (we have suppressed the $r,\theta$ arguments as before). Then, at the bend entry point $z=0$, the bent fiber wave function $\phi_{l,m}^{+(b)}(z)$ must reduce to the straight fiber wave function, $\phi^+_{l,m}(z)$ (obtained from Eq. 1 by replacing $O_{l,m}$ with $O^+_{l,m}$). Setting $z=0$, we then have
\begin{equation}
\phi_{l,m}^{+(b)}(0)=\phi^+_{l,m}(0) = O^+_{l,m}.
\end{equation}
We can now express $O^+_{l,m}$ in terms of the perturbed amplitudes, $O^{'+}_{l',m'}$ of the bent fiber (see Sections   3 and  4.4), 
\begin{equation}
O^+_{l,m}=\sum_{l',m'}{\alpha^+_{(l,m)(l',m')}O^{'+}_{l',m'}}.
\end{equation}
 Assuming orthonormality of $O'_{l,m}$'s, 
\begin{equation}
\alpha^{+}_{(l,m),(l',m')}= <O^{'+}_{l',m'}|O^{+}_{l,m}>.
\end{equation}
Inserting Eq. 30 into Eq. 33, we obtain
\begin{equation}
\alpha^{+}_{(l,m)(l',m')}= \delta_{ll'}\delta_{mm'}+c_{(l',m')(l,m)} (l'\ne\pm  l),
\end{equation}
where $c_{(l',m')(l,m)}$ encapsulates contributions in all orders of $\lambda'$:
\begin{equation}
c_{(l',m')(l,m)}=a^{(1)}_{(l',m')(l,m)}+ a^{(2)}_{(l',m')(l,m)}+... (l' \ne \pm l).
\end{equation}
At the end of the bend ($z=L$), from Eqs. 31 and 32, we have
\begin{equation}
\phi_{l,m}^{+(b)}(L)=\sum_{l',m'}{\alpha^+_{(l,m)(l',m')}O^{'+}_{l',m'}}e^{i\beta^{'+}_{l',m'}L},
\end{equation}
where $\beta^{'+}_{l',m'}$ is the propagation constant associated with the $O^{'+}_{l',m'}$ perturbed mode (see Section 4.4). Substituting Eq. 34 and 30  in the above expression, we obtain
\begin{equation}
\begin{split}
\phi_{l,m}^{+(b)}(L)=O^+_{l,m}e^{i\beta^{'+}_{l,m}L}&+\sum'_{l',m'}c_{(l,m)(l',m')}O^+_{l',m'}e^{i\beta^{'+}_{l,m}L}+\sum'_{l',m'}c_{(l',m')(l,m)}O^+_{l',m'}e^{i\beta^{'+}_{l',m'}L}\\ &+\sum'_{l',m'}\sum'_{l",m"}c_{(l',m')(l,m)}c_{(l',m')(l",m")}O^+_{l",m"}e^{i\beta^{'+}_{l',m'}L},
\end{split}
\end{equation}
where the primed summations exclude the $ l$ values.
Except for normalization, Eq. 37 then is the complete solution to all orders in perturbation for the incident $OAM^+_{l,m}$ mode which exits the bend at $z=L$ (see Fig. 2). For an incident $OAM_{l,m}^-$ mode, we obtain an identical expression, except that $-$ superscript replaces the $+$ superscript everywhere.
For an incident  $OAM_{l,m}$ mode, in which we are interested, we combine the above two results in   accordance with Eq. 24   as follows:
\begin{equation}
\phi_{l,m}^{(b)}( L)=\frac{1}{\sqrt{2}}\Big(\phi_{l,m}^{+(b)}(L)+\phi_{l,m}^{-(b)}( L)\Big).
\end{equation}
  If we set the $c$ coefficients above to zero (see Eqs. 34 and 35), effectively meaning there is \emph{no interaction} with the other modes, $OAM_{l',m'}, l'\ne \pm l$, Eq. 38, utilizing Eq. 37, immediately gives back the result (Eqs. 25 and 26) we had obtained earlier in Section 4.3. 
 For the $l=0$ incident mode (the nondegenerate case), we  get rid of the $+$ superscript everywhere in Eq. 37, obtaining in the process an expression in terms of the  conventional OAM mode amplitudes, $O_{n,k}$. Note that $\beta'_{0,m}=\beta_{0,m}$ in first order, but becomes different from $\beta_{0,m}$  in second order (see Eqs. 17 and 18).
\begin{footnotesize}
\begin{figure}[htbp]
  \centering
  \includegraphics[width=11cm]{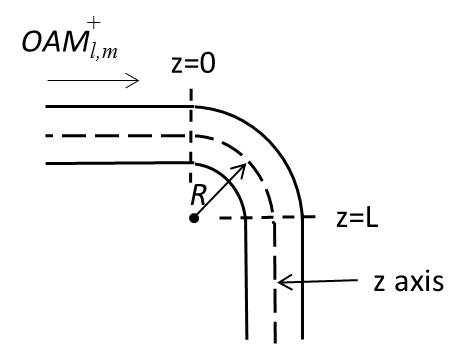}
\caption {An $OAM^+_{l,m}$ mode   (with a circular polarization)  in a straight fiber with an amplitude $O^+_{l,m}$ encountering a bend of radius $R$ and length $L$; $OAM^+_{l,m}$ mode is a linear combination of the two degenerate OAM modes, $OAM_{l,m}$ and $OAM_{-l,m}$ with amplitudes, $O_{l,m}$ and $O_{-l,m}$, respectively; $O^+_{l,m}=\frac{1}{\sqrt{2}}(O_{l,m}+O_{-l,m})$,   which corresponds to the field of the $LP_{l,m}^{(e)}$ mode (see Section 4.3).}
\end{figure}
\end{footnotesize}
\subsection{First Order Solution}
Extracting  terms up to first order   in $\lambda'$, which originate in the first three terms on the RHS of Eq. 37, we obtain
\begin{equation}
\phi_{l,m}^{+(b)}( L)=O^+_{l,m}e^{i\beta^{'+}_{l,m}L}+\sum_{l'=l\pm 1,m'}a^{(1)}_{(l,m)(l',m')}O^+_{l',m'}e^{i\beta^{'+}_{l,m}L}-\sum_{l'=l\pm 1,m'}a^{(1)}_{(l,m)(l',m')}O^{+}_{l',m'}e^{i\beta^{'+}_{l',m'}L},
\end{equation}
where we have used the fact that $a^{(1)}_{(l',m')(l,m)}=-a^{(1)}_{(l,m)(l',m')}$; the summations include only $l'=l\pm 1$. 
Similarly, for the incident mode $OAM^-_{l,m}$, we have
\begin{equation}
\phi_{l,m}^{-(b)}( L)=O^-_{l,m}e^{i\beta^{'-}_{l,m}L}+\sum_{l'=l\pm 1,m'}a^{(1)}_{(l,m)(l',m')}O^-_{l',m'}e^{i\beta^{'-}_{l,m}L}-\sum_{l'=l\pm 1,m'} a^{(1)}_{(l,m)(l',m')}O^-_{l',m'}e^{i\beta^{'-}_{l',m'}L}.
\end{equation}
Combining the above two results in accordance with Eq. 38 yields the result for the incident $OAM_{l,m}$ mode: 
\begin{equation}
\begin{split}
\phi_{l,m}^{(b)}( L)&=(\cos(\Delta\beta'_{l,m}L/2)O_{l,m}+ i \sin(\Delta\beta'_{l,m}L/2) O_{-l,m})e^{i\beta_{l,m}L}\\
& +2i\sum_{l'=l\pm l,m'} a_{(l,m)(l',m')}^{(1)}(\cos(\Delta\beta'_{l',m'}L/2)O_{l',m'}+ i \sin(\Delta\beta'_{l',m'}L/2) O_{-l',m'})(\sin(\beta_{l,m}-\beta_{l',m'})L/2) e^{i(\beta_{l,m}+\beta_{l',m'})L/2};
\end{split}
\end{equation}
  $\Delta\beta'_{l',m'}$ is calculated from Eqs. 28 and 22 and related to the $OAM_{l',m'}$ mode's $2\pi$ walk-off length via Eq. 27 after replacing $l,m$ with $l',m'$ everywhere.  
The first term on the right-hand-side of the above equation is the same as the one given in Section 4.3 (see Eq. 26), where we completely ignored the interactions with the other modes. The interaction  (mixing)  with the other modes is now captured in the second part of the right-hand-side where the  $l'$ modes mixed (in first order) also display an oscillating behavior   with their negative $l'$ counterpart on account of the bend effect.   This oscillating behavior is, however, relatively weak as it is  suppressed by the first order perturbation mixing coefficient,   $a_{(l,m)(l',m')}^{(1)}$.  An additional multiplicative sine factor in Eq. 41 is a consequence of the interference between the input $OAM_{l,m}$ mode and the $OAM_{l',m'}$ mode. The  impact of this interference (the sine factor value)   depends upon the modes'  propagation constant difference and the length $L$ traversed within the bend. Because this propagation constant difference is normally much larger than the $\Delta \beta'_{l,m}$ difference, the sine factor is very sensitive to the change in $L$ value.  We note here that for  the  $l=0$  nondegenerate case, we can set $\Delta\beta'_{l,m}=0$ in Eq. 41. We also note here that due to the orthogonality of $O_{l,m}$'s, to first order in $\lambda'$ (i.e., neglecting second order terms in $\lambda'$), $<\phi_{l,m}^{(b)}( L)|\phi_{l,m}^{(b)}( L)>=<\phi_{l,m}^{(b)}( 0)|\phi_{l,m}^{(b)}( 0)>$.    In other words, the total output power is the same as the input power. We also verify from Eq. 41 that for $L=0$, $\phi_{l,m}^{(b)}( L)=O_{l,m}$,  the input amplitude, as one would expect.
\\\\
$\phi_{l,m}^{(b)}(z\ge L)$, which corresponds to propagation within the straight portion of the fiber beyond $z=L$ (see Figure 2), is
obtained by replacing the first exponential in Eq. 41, $e^{i\beta_{l,m}L}$, with $e^{i\beta_{l,m}z}$, and the second exponential, $e^{i(\beta_{l,m}+\beta_{l',m'})L/2}$, with $e^{i(\beta_{l,m}-\beta_{l',m'})L/2}e^{i\beta_{l',m'}z}$.   We see at once that the obtained expression reduces to the expression, Eq. 42, at $z=L$, verifying the validity of the foregoing substitutions. At $z>L$, within the straight fiber, apart from the common propagation phase factors, $e^{i\beta_{l,m}z}$ for the $OAM_{l,m}$ and $OAM_{-l,m}$ modes, and $e^{i\beta_{l',m'}z}$ for the $OAM_{l',m'}$ and $OAM_{-l',m'}$ mode pair, the magnitudes of the amplitudes of the component modes within the mixture remain fixed (frozen) at values determined at the bend output, $z=L$. 
\\\\
Summarizing,  the input $OAM_{l,m}$ after entering the bend, suffers transformations into its negative $l$ counterpart,  mixing also with the neighboring modes (determined by the $|\Delta l|=1$ rule), which, in turn, also transform into their corresponding negative counterparts.   These transformations occur continuously throughout the bend until the mode mixture emerges at $z=L$, with a mode composition profile described by Eq. 41. Thereafter ($z>L$),  except for the  phase changes that occur as described above, the mode mixture propagates with the magnitudes of the individual  amplitudes (determined at $z=L)$ intact.

\subsection{Second Order Solution}
From Eq. 37  we determine the second order contribution to be 
\begin{equation}
 \sum_{l'=l\pm 2,m'}(a^{(2)}_{(l,m)(l',m')}O^+_{l',m'}e^{i\beta^{'+}_{l,m}L}+a^{(2)}_{(l',m')(l,m)}O^+_{l',m'}e^{i\beta^{'+}_{l',m'}L})\\
 -\sum_{l'=l\pm 1,m'}a^{(1)}_{(l,m)(l',m')}\sum_{l"=l'\pm 1,m"}a^{(1)}_{(l',m')(l",m")}O^{+}_{l",m"}e^{i\beta^{'+}_{l",m"}L}.
\end{equation}
 However, an additional term  quadratic in $\lambda'$,  arising from a proper normalization of series expansion up to second order in $\lambda'$  (see Eq. 30 or Eq. 6), will also have to be added to the above expression for completeness. A similar expression with the same additional term when the incident mode is $OAM^-_{l,m}$ follows. Combining the two contributions in accordance with Eq. 38 then yields the second order contribution for an input $OAM_{l,m}$ mode, with the resultant oscillatory behavior damped by the square of the perturbation parameter $\lambda'$.  As a result, we do not concern ourselves with the detailed analytic expressions here. 

\subsection{Crosstalk}
  The orthogonality of $OAM_{l,m}$ modes allows for simultaneous transmission of these modes  in a spatial (mode)-division multiplexed  fiber system. So it is critical to keep as small as possible the amount of mode mixing in order to avoid any significant crosstalk.  From Eq. 41,  we can calculate crosstalk (or charge weight \cite{yue, wang}), $X_{(l,m)(l',m')}$ for the various component $OAM_{l',m'}$ modes of the output, $\phi^{(b)}_{l,m}(L)$, as 
\begin{equation}
X_{(l,m)(l',m')} = 10 log_{10}|<O_{l',m'}|\phi^{(b)}_{l,m}(L)>|^2,
\end{equation}
where  $X_{(l,m)(l',m')}$ is expressed in dB. 
\\\\
Invoking the orthogonality of the OAM modes and using Eqs. 41 and 27, we immediately see that $X_{(l,m)(l',m')}$ is equal to $\cos^2(\pi L/L^{(2\pi)}_{l.m})$ for $l'=l$ and  equal to $\sin^2(\pi L/L^{(2\pi)}_{l.m})$ for $l'=-l$. Similarly, when $l'=l\pm1$ (corresponding to neighboring modes), $X_{(l,m)(l',m')} = 4|a_{(l,m)(l',m')}^{(1)}|^2 \cos^2(\pi L/L^{(2\pi)}_{l'.m'})\sin^2((\beta_{l,m}-\beta_{l',m'})L/2)$, and for $l'=-(l\pm1)$, the degenerate partners of the neighboring modes, $ X_{(l,m)(l',m')} = 4|a_{(l,m)(-l',m')}^{(1)}|^2 \sin^2(\pi L/L^{(2\pi)}_{-l',m'})\sin^2((\beta_{l,m}-\beta_{-l',m'})L/2)$; the minus signs in front of $l'$ on the RHS negate the minus sign of $l'$.  $X_{(l,m)(l',m')}$ is an explicit function of $L$, as we would expect. The latter two expressions imply a maximum possible crosstalk  given essentially by $4|a_{(l,m)(l\pm1,m')}^{(1)}|^2$. This is due to the fact that $\sin ((\beta_{l,m}-\beta_{l',m'})L/2) $ is a very rapidly varying function compared to  the sinusoidal functions involving the $2\pi$ walk-off length because $\beta_{l,m}-\beta_{l',m'}$, the propagation constant difference between the two modes, is much greater than $(2\pi)/L^{(2\pi)}_{l',m'}$ (see the discussion following Eq. 41 and Eqs. 29 and 22). 
\section{Application to Step-Index Fibers and Results}
For applications, we consider a few-mode fiber and a conventional multimode fiber. For  numerical simulations we take as  input an $OAM_{l,m}$ mode with $m=1$. We perform all calculations in MATLAB.

\subsection{Few Mode Fiber}
We assume parameters: $a=10 \mu m, n_1=1.45205, n_2=1.44681$, which approximately correspond to an actual \emph{OFS}-manufactured fiber. Then, for a wavelength $\lambda=1.55 \mu m$, we find that the normalized frequency $V=2\pi a\sqrt{(n_1^2-n_2^2)}/\lambda =4.996$. Consequently, the fiber can support only six modes: $OAM_{0,1}, OAM_{0,2}$ and the two degenerate pairs: $OAM_{1,1}, OAM_{-1,1}, OAM_{2,1}$, and $OAM_{-2,1}$. 
\subsubsection{Degeneracy breaking and the $2\pi$ walk-off length}
Here we consider the breaking of the degeneracy within the separate subspaces of the $l=1$ and $l=2$ degenerate OAM pairs.  From Eq. 22, for $l=1$, we have
\begin{equation}
\Delta\beta'^2_{1,1}= 2\lambda'^2\Big(\frac{(\delta H_{(1,1)(0,1)})^2}{\beta_{1,1}^2-\beta_{0,1}^2}+\frac{(\delta H_{(1,1)(0,2)})^2}{\beta_{1,1}^2-\beta_{0,2}^2}\Big)
\end{equation}
and for $l=2$,
\begin{equation}
\Delta\beta'^2_{2,1}= 2\lambda'^4\Big(\frac{(\delta H_{(2,1)(1,1)})^2(\delta H_{(1,1)(0,1)})^2}{(\beta_{2,1}^2-\beta_{1,1}^2)^2(\beta_{2,1}^2-\beta_{0,1}^2)}+\frac{(\delta H_{(2,1)(1,1)})^2(\delta H_{(1,1)(0,2)})^2}{(\beta_{2,1}^2-\beta_{1,1}^2)^2(\beta_{2,1}^2-\beta_{0,2}^2)}\Big).
\end{equation}
\begin{table}
\centering
\caption{The $2\pi$ walk-off length, $L^{(2\pi)}_{l,m}$, specified in meters, as a function of the bend radius $R$ for different input modes, $OAM_{l,m}$; it varies as $R^{2l}$ for fixed $l,m$.}
\begin{tabular}{ |c|c|c|c|} 
 \hline
$ l,m$& R=4cm & R=8cm&R=16 cm \\ 
\hline
 1,1 & 0.146& 0.585&2.34 \\ 
\hline
 2,1&416& 6.65 x$10^3$ & 1.06 x$10^5$ \\ 
 \hline
\end{tabular}
\end{table}
The $\delta H$ matrix elements are numerically calculated in accordance with Eq. 12.
In Table I, we give the results for the $2\pi$ walk-off length, $L^{(2\pi)}_{l,m}$, for different values of the bend radius $R$.  $L^{(2\pi)}_{l,m}$ increases as $R^2$ for the $l=1$ mode and as $R^4$ for the $l=2$ mode, becoming extremely large for high values of $R$. Going from an $l$ state to a $-l$ state involves an OAM change of $2l$. The fact that the $2\pi$ walk-off length is much larger for $l=2$ as compared to $l=1$ is due to the fact that it becomes increasingly difficult for the given bend to impart an OAM change of 4 units  as opposed to 2 units (see Section 4.3).  

\subsubsection{Mixing with Neighboring Modes}
\underline{Input $OAM_{1,1}$ Mode}
\\\\
Due to the selection rule: $\Delta l=\pm1$, $OAM_{1,1}$ can mix with $OAM_{0,1}, OAM_{0,2}$, and $OAM_{2,1}$ in first order, and with $OAM_{-2,1}$ (indirectly) via the mixing of $OAM_{2,1}$ with $OAM_{-2,1}$ (see   Section 6.1.1).   For $R=4 cm$, the coefficients, $a^{(1)}_{(1,1)(0,1)}, a^{(1)}_{(1,1)(0,2)}$, and $a^{(1)}_{(1,1)(2,1)}$ are respectively found to be $-0.066321, -0.009947$, and $0.03488$. $\Delta \beta'_{1,1}$ and $\Delta\beta'_{2,1}$, which determine the  $2 \pi$ walk-off lengths for the corresponding modes, are respectively $-42.97/m$ and $-0.01511/m$. The contributions of the various   $OAM$ modes, including the original input   $OAM_{1,1}$ mode, to the final state (see Eqs. 41  and 27)  are
\\\\
1)$|<O_{1,1}|\phi^{(b)}_{1,1}(L)>|=|\cos(\pi L/L_{(1,1)}^{(2\pi)})|=|\cos(43L/2)|$.
\\
2)$|<O_{-1,1}|\phi^{(b)}_{1,1}(L)>|=|\sin(\pi L/L_{(1,1)}^{2\pi)})|=|\sin(43L/2)|$.
\\
3)$|<O_{0,1}|\phi^{(b)}_{1,1}(L)>|=2| a^{(1)}_{(1,1)(0,1)}\sin((\beta_{1,1}-\beta_{0,1})L/2)|=1.33x10^{-1}|\sin(2.534x10^3L)|$.
\\
4) $|<O_{0,2}|\phi^{(b)}_{1,1}(L)>|=2| a^{(1)}_{(1,1)(0,2)}\sin((\beta_{1,1}-\beta_{0,2})L/2)=1.99x10^{-2}|\sin(4.115x10^3L)|$.
\\
5)$|<O_{2,1}|\phi^{(b)}_{1,1}(L)>|=2| a^{(1)}_{(1,1)(2,1)}\cos(\pi L/L_{(2,1)}^{(2\pi)})(\sin((\beta_{1,1}-\beta_{2,1})L/2)|
\\
=7.0x10^{-2}|\cos(7.5x10^{-3}L)\sin(3.2x10^3L)|$.
\\
6)$|<O_{-2,1}|\phi^{(b)}_{1,1}(L)>|=2| a^{(1)}_{(1,1)(2,1)}\sin^2(\pi L/L_{(2,1)}^{(2\pi)})(\sin^2((\beta_{1,1}-\beta_{2,1})L/2)|
\\
=7.0x10^{-2}|\sin(7.5x10^{-3}L)\sin(3.2x10^3L)|$.
\begin{footnotesize}
\begin{figure}[htbp]
  \centering
  \includegraphics[width=11cm]{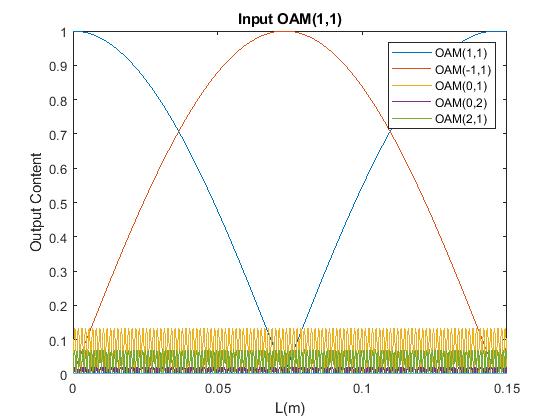}
\caption {Various   $OAM_{n,k}$ contributions to the  output state, $\phi^{(b)}_{1,1}$, as a function of the bend length, $L$   (see text for more details); the input is $OAM_{1,1}$; a bend radius $R=4 cm$ is assumed; while the dominant modes, the $OAM_{11}$ and its admixed negative $l$ counterpart, $OAM_{-1,1}$, stand out, the rest - $OAM_{0,1}$ (orange-yellow), $OAM_{0,2}$ (purple), $OAM_{2,1}, $ (green), and $OAM_{-2.1}$ (extremely small and not visible)- form a rapidly varying clutter that varies inversely with the bend radius $R$, and fades away as $R$ increases.}
\end{figure}
\end{footnotesize}
\\
We see from above that the admixed neighboring modes' contributions (3-6) are extremely small.  As a result, the oscillating modes, $OAM_{1,1}$ and $OAM_{-1,1}$ stand out.  Furthermore, the contributions of these admixed modes are in inverse proportion to  the radius of the bend $R$, and hence decrease further with increasing radius $R$. Figure 3 shows the plot of the contributions of the various admixed modes, including the dominant oscillating $OAM_{1,1}$ and $OAM_{-1,1}$ modes. The small oscillations are due to the large propagation constant difference multiplying the $L$ variable in the argument of the sine functions. The admixed $OAM_{-2,1}$ mode derives its  amplitude from the coupling of the $OAM_{1,1}$ mode to the $OAM_{2,1}$ mode in first order, and the subsequent $OAM_{2,1}$ transformation into $OAM_{-2,1}$. Thus, for distances $L<<L^{(2\pi)}_{2,1}$, it is negligibly small, and thus not visible in Fig 3.  One notes here that  no second order contributions are possible as $l=1$ can connect with $l=0$ and $l=2$ in first order only, and there are no modes with $l$ greater than 2.
\\\\
\underline{Input $OAM_{2,1}$ Mode}
\\\\
Invoking Eq. 41, we find  for the input $OAM_{2,1}$
\begin{equation}
\begin{split}
\phi_{2,1}^{(b)}( L)&=(\cos(\Delta\beta'_{2,1}L/2)O_{2,1}+ i \sin(\Delta\beta'_{2,1}L/2) O_{-2,1})e^{i\beta_{2,1}L}\\
& +2i a_{(2,1)(1,1)}^{(1)}(\cos(\Delta\beta'_{1,1}L/2)O_{1,1}+ i \sin(\Delta\beta'_{1,1}L/2) O_{-1,1})(\sin(\beta_{2,1}-\beta_{1,1})L/2) e^{i(\beta_{2,1}+\beta_{1,1})L/2}.
\end{split}
\end{equation}
There is mixing with $OAM_{1,1}$ (due to the selection rule $\Delta l=\pm1$), and   with  $OAM_{-1,1}$ through the breaking of degeneracy between $OAM_{1,1}$ and $OAM_{-1,1}$. The first order mixing given by $a^{(1)}_{(2,1)(1,1)}=-a^{(1)}_{(1,1)(2,1)}=-0.03488$. $\Delta \beta'_{2,1}=-0.01511/m$ and $\Delta\beta'_{1,1}=-42.97/m$. By taking the appropriate scalar products, the contributions of the modes comprising the final state are determined and plotted in Fig. 4: 
\\\\
1)$|<O_{2,1}|\phi^{(b)}_{2,1}(L)>|=\cos(\pi L/L^{(2\pi)}_{2,1}|=\cos(0.0075L)$.
\\
2)$|<O_{-2,1}|\phi^{(b)}_{2,1}(L)>|=|\sin(\pi L/L^{(2\pi)}_{2,1}|=|\sin(0.0075L)|$.
\\
3)$|<O_{1,1}|\phi^{(b)}_{2,1}(L)>|=2| a^{(1)}_{(2,1)(1,1)}\cos(\pi L/L^{(2\pi)}_{1,1})(\sin(\beta_{2,1}-\beta_{1,1})L/2)|=0.0698|\cos(21.5L)\sin(3200L)|$.
\\
4)$|<O_{-1,1}|\phi^{(b)}_{2,1}(L)>|=2| a^{(1)}_{(2,1)(1,1)}\sin(\pi L/L^{(2\pi)}_{1,1})(\sin(\beta_{2,1}-\beta_{1,1})L/2)|=0.0698|\sin(21.5L)\sin(3200L)|$.
\\\\
  In the above expressions, we have utilized the relationship, Eq. 27. 
\begin{footnotesize}
\begin{figure}[htbp]
  \centering
 \begin{subfigure}[b]{0.4\linewidth}
    \includegraphics[width=\linewidth]{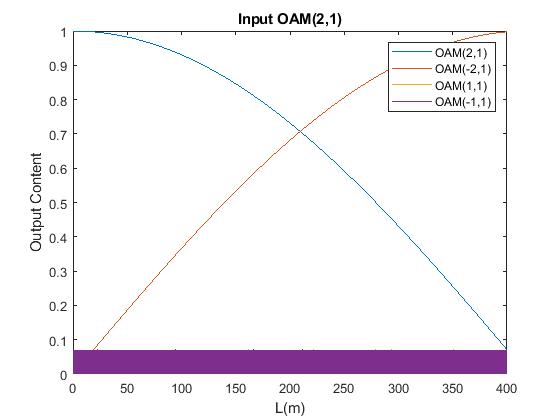}
 \end{subfigure}
  \begin{subfigure}[b]{0.4\linewidth}
    \includegraphics[width=\linewidth]{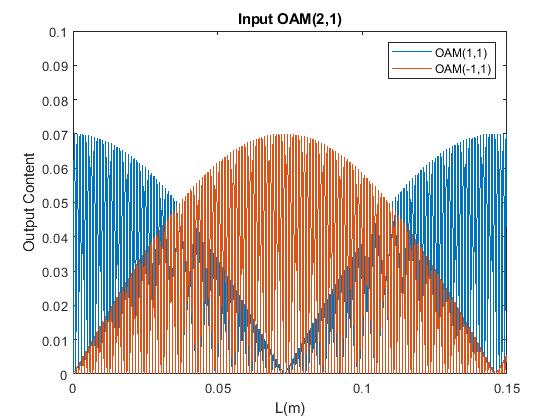}
 \end{subfigure}
  \label{fig:coffee}
 \caption {a) Various   $OAM$ mode contributions to  the output state, $\phi^{(b)}_{2,1}$, as a function of the bend length, $L$   (see text for more details); the input mode is $OAM_{2,1}$; a bend radius $R=4 cm$ is assumed; while the $OAM_{2,1}$ and $OAM_{-2,1}$ amplitudes are clearly displayed, the smaller $OAM_{1,1}$ and $OAM_{-1,1}$ amplitudes are bunched up to give the appearance of a band due to their rapid sinusoidal behavior; this band-forming clutter varies inversely with fiber bend radius $R$; b) Same as Fig. 4a, except the abscissa, ordinate ranges are reduced to bring out the detailed behavior of the admixed $OAM_{1,1}$ and the $OAM_{-1,1}$ modes; one now sees the modes' $2\pi$ walk-off-length oscillating behavior modulated by the rapid sinusoidal variations.}
\end{figure}
\end{footnotesize}
\\\\
\emph{Second Order Contributions}
\\\\
In addition to the square of the first order mixing coefficients, $a^{(1)}_{(l,m)(l',m')}$ (see Eq. 42), there will be second order coefficients that connect the $l=2$ state to the $l=0$ state via two successive applications of the $\Delta l=\pm1$ rule:
\begin{equation}
a^{(2)}_{(2,1)(0,1)}= \lambda'^2(\frac{(\delta H_{(0,1)(1,1)})(\delta H_{(1,1)(2,1)})}{(\beta_{2,1}^2-\beta_{1,1}^2)(\beta_{2,1}^2-\beta_{0,1}^2)},
\end{equation}
\begin{equation}
a^{(2)}_{(2,1)(0,2)}= \lambda'^2(\frac{(\delta H_{(0,2)(1,1)})(\delta H_{(1,1)(2,1)})}{(\beta_{2,1}^2-\beta_{1,1}^2)(\beta_{2,1}^2-\beta_{0,2}^2)}).
\end{equation}
 Substitution of the various parameter values in the above expression yields  numerical values of $1.023x10^{-3}$ and $1.561x10^{-3}$ for the above coefficients, respectively. These are much smaller than the mixing coefficients, $ a^{(1)}_{(2,1)(\pm1,1)}$, calculated in first order,   and not considered further.
\subsubsection{Impact of Crosstalk in Mode-Multiplexed Few Mode Fiber Systems}
Because different $OAM_{l,m}$ modes by virtue of their orthogonality can travel simultaneously on the same fiber, it is important to ensure that a fiber bend does not lead to an unacceptably large crosstalk due to mode-mixing. Crosstalk (in dB) between a given input mode $OAM_{l,m}$ and other component modes, $OAM_{l',m'}$,  of the output mode mixture, $\phi^{(b)}_{l,m}(L)$, was defined in Section 5.3.  In practical scenarios,  fiber bends may manifest themselves as some  loose fiber coiled up  between the input and output of a fiber transmission system, or as some fiber wound up on a spool as happens in many experimental setups (the fiber bend  length  $L= 2\pi RN$, where $N$ is the number of turns within the fiber coil/spool). Table 2 illustrates the crosstalk (in dB) for the input $OAM_{2,1}$ as a function of the bend length $L$.  The radius $R$ of the fiber bend  is fixed at $4cm$, the considered $R$ value in the mixing mode calculations in Section 6.1.2. We see from the table that the  $OAM_{2,1}$ mode content within the output is essentially unchanged for $L=2m$ and $L=10m$, because  $L<< L_{2,1}^{(2\pi)} (=416 m)$. The crosstalk consequently with $OAM_{-2,1}$ is very small. This crosstalk, however, increases with $L$, and at $L=100m$, which roughly corresponds to $L=L_{2,1}^{(2\pi)}/4$, $OAM_{-2,1}$ appears with approximately the same strength (dB value) as $OAM_{2,1}$. At $L=200m\approx L_{2,1}^{(2\pi)}/2$, the full conversion of $OAM_{2,1}$ to $OAM_{-2,1}$  (and vice versa) has occurred as indicated by the crosstalk values, and finally, at $L=400m\approx L_{2,1}^{(2\pi)}$, the $OAM_{2,1}$ mode has recovered itself with very small crosstalk with its negative $l$ counterpart, the $OAM_{-2,1}$ mode. During this entire cycle, i.e., for all the $L$ values considered in Table 2, the crosstalk with $OAM_{1,1}$ and $OAM_{-1,1}$ modes remains consistently low due to a factor of $\lambda'^2$, which appears in the modulus square of the coefficient $a^{(1)}_{(2,1)(1,1)}$ (see Section 5.3). Consequently, the output intensity pattern during this cycle will basically change from a donut shape to the four lobe $LP_{2,1}$ pattern tilted by $\pi/8$ and then back to a (nearly) donut shape at $L=400m$ (see Section 4.3). 
\\\\
The maximum mode content of $OAM_{1,1}$ or $OAM_{-1,1}$  is 0.0698 (see calculations following Eq. 46 as well as Figs. 4a and b); this value translates to a maximum possible crosstalk of -23 dB for each of these modes. If we now set the criterion that  crosstalk with each mode ($l' \ne 2$) is to not exceed -23 dB, we must ensure that $OAM_{-2,1}$ mode also remains below the threshold of $-23 dB.$ Since the crosstalk with $OAM_{-2,1}$ mode is a function of $L$,   then the maximum $L$ permitted, which we denote by $L_{max}$, is obtained by setting the magnitude of the admixed $OAM_{-2,1}$ mode to 0.0698. In other words, we set $|\sin(\pi L_{max}/L_{2,1}^{(2\pi)})|=0.0698$. Confining the analysis to  $0\le L \le L_{2,1}^{(2\pi)}$, this yields $L_{max}=2.22x 10^{-2}L_{2,1}^{(2\pi)}=9.24 m$. 
\begin{table}
\centering
\caption{  Crosstalk, $X_{(2,1)(l',m')}$ (in dB) for the various component $OAM_{l',m'}$ modes within the $OAM_{2,1}$ output mode mixture as a function of $L$; the fiber bend radius $R$ is fixed at 4 cm. The $2\pi$ walk-off length $L_{2,1}^{(2\pi)} = 416m$ (see Table 1).}
\begin{tabular}{ |c|c|c|c|c|c|} 
 \hline
$ l',m'$& L=2m & L=10m & L=100m & L=200m & L=400m\\ 
\hline
 2,1 & -.001&-.025&-2.71&-24.4 &-.064\\ 
\hline
 -2,1&-36.4& -22.4 &-3.33& -.016 & -18.4\\ 
 \hline
 1,1&-33.5&-48.9 &-37.1& -28.0 & -43.4\\ 
 \hline
 -1,1&-30.0&-35.0 &-30.1& -27.0 & -24.2\\ 
 \hline
\end{tabular}
\end{table}
\\\\
 Noting $L_{2,1}^{(2\pi)}\propto R^{4}$ and $a_{(2,1)(1,1)}^{(1)}\propto R^{-1}$ , similar crosstalk tables can then be constructed and analyzed for any  radius $R$ of the fiber coil that may exist between the input and the output of an $OAM$ transmission system. For example, for $R=10.5cm$ (approximately the radius of a  commercial fiber spool),  $L_{2,1}^{(2\pi)}$ will increase from $416m$ to $(10.5/4)^4 x 416 m = 19.75km$.  Using the previous criterion of  $L_{max}=2.22x 10^{-2} L_{2,1}^{(2\pi)}$ (based on crosstalk not exceeding -23 dB), $L_{max}$ is calculated to be 438.45m, implying a coil with a radius $R\ge10.5cm$  and coil length $L\le 438.45 m$ has practically no mode-mixing impact on $OAM_{2,1}$ transmission. Exactly, the same result holds for the transmission of $OAM_{-2,1}$ due to the (almost) identical form of the mode-mixing analytic expressions for the negative $l$ topological charge. 
\\\\
A similar calculation and analysis for $OAM_{1,1}$ transmission can also be performed for a variety of $R$ and $L$ values and the results tabulated. Because the $2\pi$ walk-off length for the $l=1$ mode is much smaller compared to the $l=2$ case (see Table 1), we expect greater crosstalk  sensitivity as a function of $L$ in the $l=1$ case. As a result, $L_{max}$, for a given value of $R$ will be much smaller here.   Thus, if OAM modes with topological charges of $1, -1, 2, -2$ were all to be multiplexed on the same fiber, the limits  imposed by the $l=1$ case  will predominate.
%
%
%

%
\subsection{Conventional Multimode Fiber}
Here the numerical work assumes $n_1=1.461, n_2= 1.444, a=25\mu m$; the parameters correspond to a real \emph{ThorLabs} multimode step-index fiber. For a wavelength of $\lambda=1.55 \mu m$, the normalized frequency $V=22.5$, implying a  maximum value of $l$ equal to $18$. There are over two hundred modes, but for illustrative purposes we consider modes ranging from $l=0$ to $l=6$. Unless otherwise stated, we take $R= 4 cm$ for which $\lambda'=a/R=6.25 x 10^{-4}$.
\\\\
 Figure 5a shows the calculated effective refractive indices, $n_{eff}$ of the modes above a cut-off value of 1.456; as a result, some of the higher radial order $m$ modes are not shown. The effective refractive index, $n_{eff}$ decreases in  magnitude with increasing $m$ for fixed $l$. The spacing between consecutive $m$ indices also increases, the increase being larger for higher $l$ value modes. There are no accidental degeneracies.  We find that the effective refractive index difference $\Delta n_{eff}$ is the smallest for the $OAM_{5,1}$ and $OAM_{0,3}$ mode pair, being equal to $0.64 x 10^{-4}$.  Any mode mixing due to this small difference can occur in perturbation order five only (for this pair of modes), and is, therefore, quickly suppressed by the fifth power of $\lambda'$.  The same is true in the  case of  the modes $OAM_{6,1}$ and $OAM_{1,3}$, 
where $\Delta n_{eff}=1.42 x 10^{-4}$.  Calculations show that  the  only  pair with a significant mixing amplitude in Figure 5a due to the proximity of their effective refractive index values is the pair, $OAM_{2,1}$ and $OAM_{0,2}$ with  $\Delta n_{eff}=1.88x10^{-4}$. Because $|\Delta l| =2$, mixing occurs in second-order;  $a^{(2)}_{(2,1)(0,2)}$ is approximately $25\%$ of its first order counterparts, $ a^{(1)}_{(2,1)(1,1)}$. As $l$ increases, the second order contributions dwindle further which can be understood from the  increasing $\Delta n_{eff}$ values for the $\Delta l=2$ neighbors as well as the smaller $\delta H$ matrix element values.  Furthermore, the mixing is found to be most  dominant  in first order perturbation when the radial mode order difference, $\Delta m=0$, i.e., when $m=1$ for the neighboring modes as well;  for the $\Delta m \ne 0$ cases, the smaller values of the associated $\delta H$  matrix elements contribute significantly in their suppression; e.g.,  $a^{(1)}_{(2,1)(1,2)}$ is approximately $6\%$ of $ a^{(1)}_{(2,1)(1,1)}$. In Fig. 5c, we show the amplitudes for mixing with the nearest neighbor modes when $\Delta m=0$. These coefficients, as discussed above, become smaller with increasing values of $l$. The values further reduce with increasing bend radius $R$ due to the inverse $R$ relationship of $\lambda'$. 
\begin{footnotesize}
\begin{figure}[htbp]
  \centering
 \begin{subfigure}[b]{0.32\linewidth}
    \includegraphics[width=\linewidth]{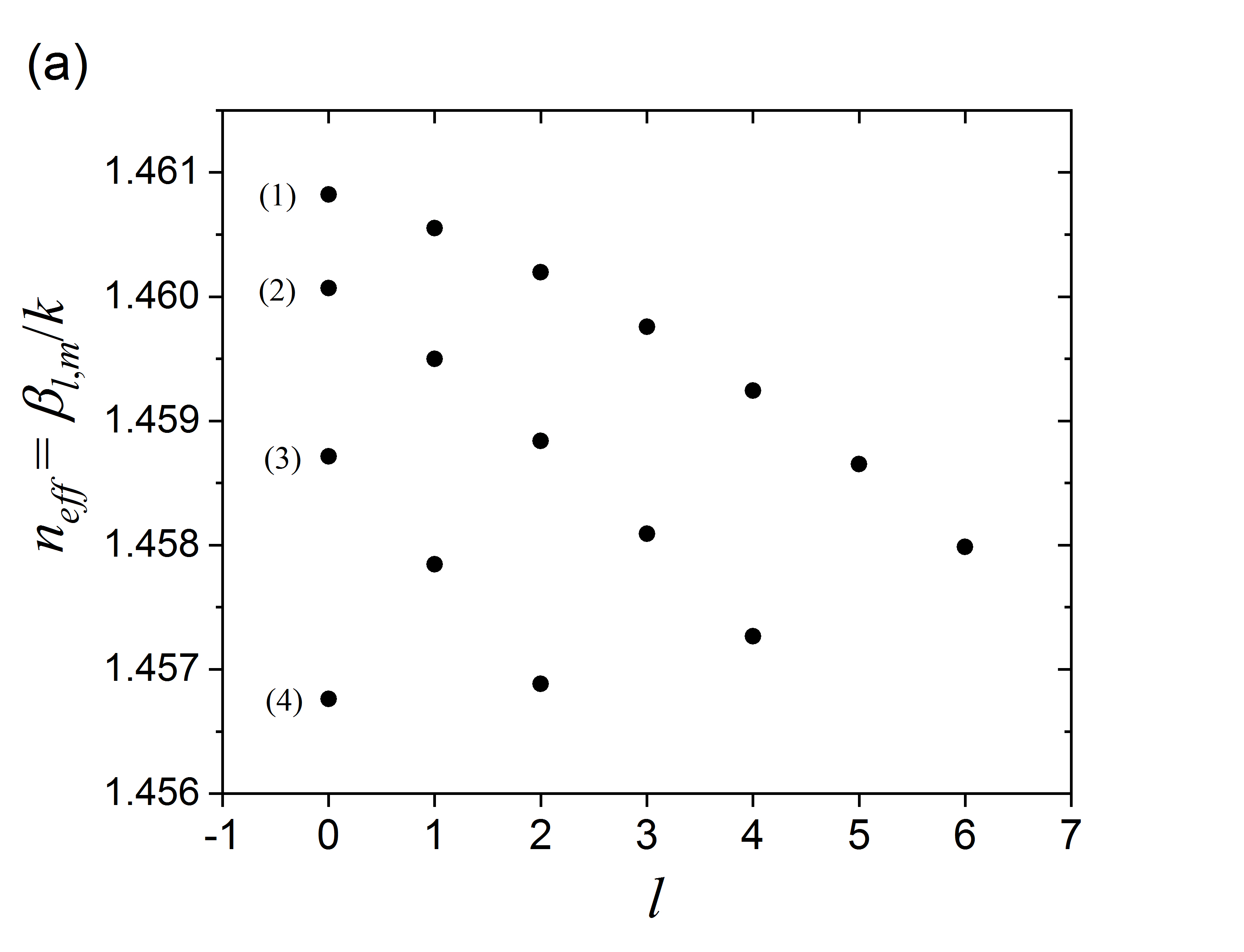}
 \end{subfigure}
 \begin{subfigure}[b]{0.32\linewidth}
    \includegraphics[width=\linewidth]{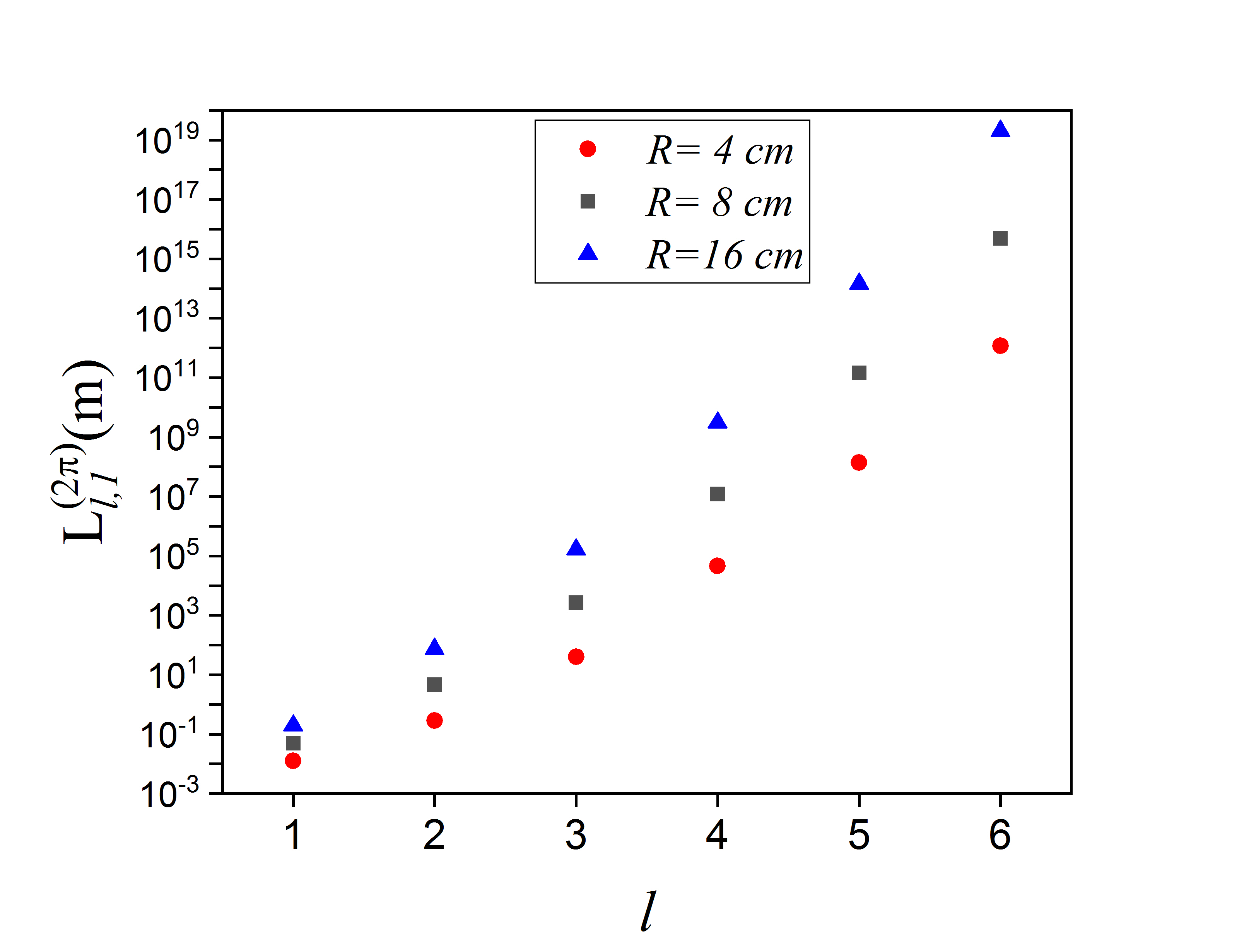}
  \end{subfigure}
 \begin{subfigure}[b]{0.32\linewidth}
    \includegraphics[width=\linewidth]{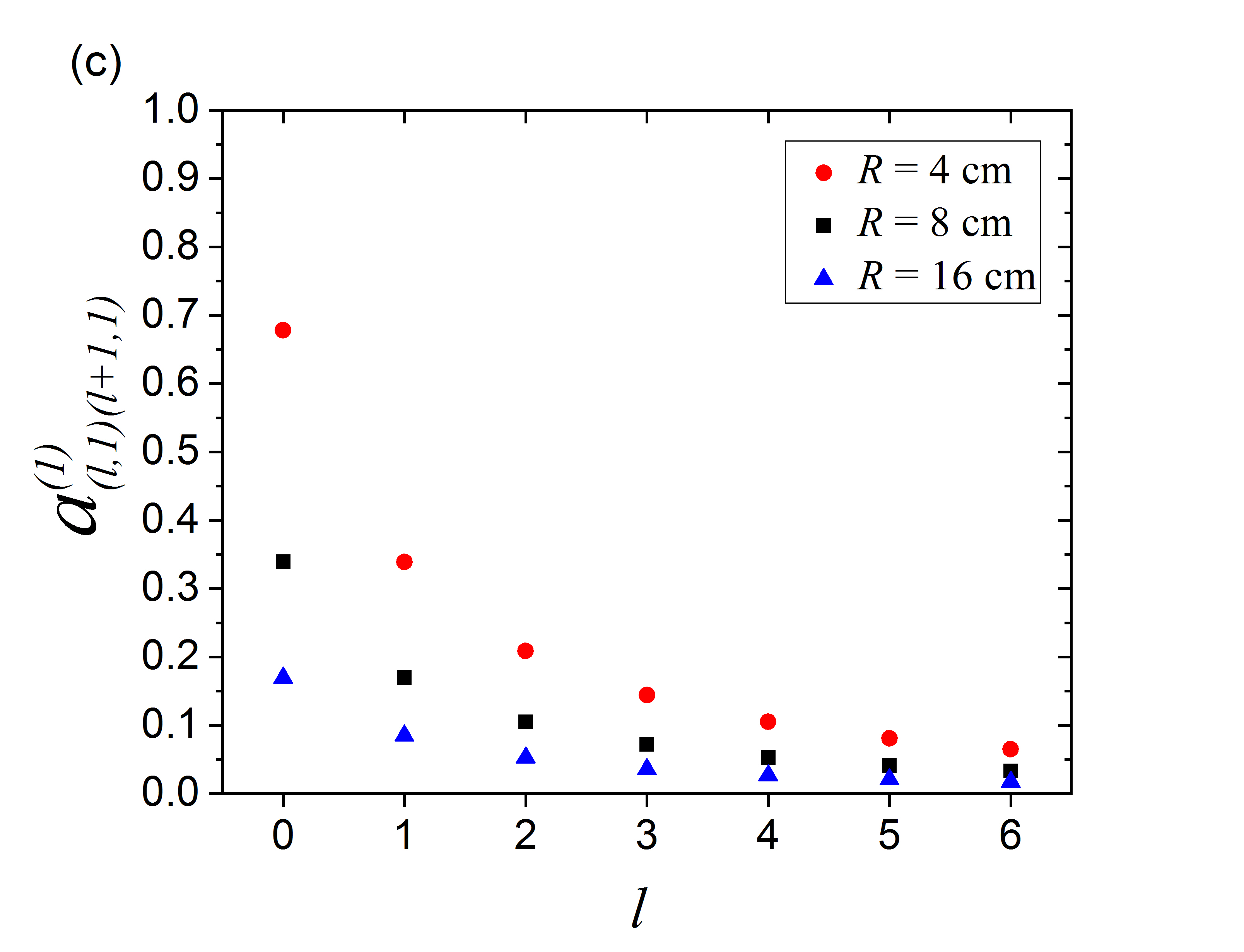}
  \end{subfigure}
  \label{multimode}
\caption{(a) Effective refractive index, $n_{eff}$ for the various modes, including higher radial order modes (indicated by values of $m$ in parentheses for the $l=0$ case, as an example). (b) $L^{(2\pi)}_{l,1}$  as a function of $l$ for different bend radii $R$. (c) The first order mixing coefficient,  $a_{(l,1)(l+1,1)}^{(1)}$, as a function of parameter $l$ for different bend radii.}
\end{figure}
\end{footnotesize}
For a fixed radius $R$, $L^{(2\pi)}_{(l,1)}$ increases sharply with $l$ (see Figure 5b), which signifies the fact that transition from $l$ OAM state to a $-l$ OAM state (or vice versa), is increasingly difficult for large $l$ values due to the requirement of transfer of $2l$ OAM (see Section 6.1); for fixed $l$, $L^{(2\pi)}_{(l,1)}$ increases with $R$ as the propagating mode experiences less curvature, as we would expect; in fact, this distance varies as $R^{2l}$, stemming from the $\lambda'^{2l}$ dependence of  $\Delta \beta^{'2}_{l,m}$ (see Eqs. 29 and 22).
\begin{table}
\centering
\caption{  a) Crosstalk, $X_{(l,m)(l',m')}$ (in dB) for the various component $OAM_{l',m'}$ modes of the $OAM_{l,1}$ output mode as a function of $L$; topological charge $l=4$ and fiber bend radius $R= 4 cm$ b) crosstalk for $l=6$ and $R=8cm$ (table on the right). } 
\begin{tabular}{ |c|c|c|c|} 
 \hline
$ l',m'$& L=100m & L=1km & L=10km\\ 
\hline
 4,1 &0&-0.02&-2.2\\ 
\hline
 -4,1&-43.4& -23.4&-4.1\\ 
 \hline
 3,1&-27.2 &-41.0& -15.4\\ 
 \hline
 -3,1&--11.0 &-10.8& -25.3\\ 
 \hline
5,1&-14.1&-24.8&-22.7\\
\hline
-5,1&-127.1&-117.7&-95.7\\
\hline
\end{tabular}
\quad
\begin{tabular}{ |c|c|c|c|} 
 \hline
$ l',m'$& L=100m & L=1km & L=10km\\ 
\hline
 6,1 &0&0&0\\ 
\hline
 -6,1&-263.9& -243.9&-223.9\\ 
 \hline
 5,1&-27.9 &-22.9& -21.9\\ 
 \hline
 -5,1&-201.0&--176.0& -155.0\\ 
 \hline
7,1&-30.1&-24.1&-33.7\\
\hline
-7,1&-294.0&-268.0&-257.6\\
\hline
\end{tabular}
\end{table}

\subsubsection{Crosstalk Impact in a Mode-Multiplexed Multimode Fiber System}

We must also assess the crosstalk due to a fiber bend and the limitations it can impose within an OAM mode-multiplexed multimode fiber system.  Because  $L^{(2\pi)}_{l,1}$  rises sharply with increasing $l$ (Figure 5b) and the first order mixing coefficient,  $a_{(l,1)(l+1,1)}^{(1)}$, decreases with $l$ (Figure 5c), it is preferable to use higher values of $l$ in $OAM$ mode transmission to minimize the mode-mixing impact of fiber bends. In Table 3a (first table on the left), we show the crosstalk results for  $l=4$ and $R=4cm$ as calculated from Eq. 43. We see the input $OAM_{4,1}$ mode slowly reducing in amplitude and changing into $OAM_{-4,1}$ mode as $L$ is increased from $100m$ to $10 km$.  There is a significant mixing of the neighboring modes with topological charge $l'=3$ or $5$ for all values of $L$. The maximum crosstalk possible with each of the $OAM_{\pm3,1}$ modes is given by  $ 10log_{10}(4|a^{(1)}_{(4,1)(3,1)}|^2) =-10.8 dB$ (see Section 5.3).  Similarly, with the $OAM_{\pm5,1}$ modes, the maximum crosstalk possible is  $ 10log_{10}(4|a^{(1)}_{(4,1)(5,1)}|^2)=-13.6 dB$. The results in Table 3a are consistent with these limits. To further reduce this crosstalk, it is evident from Fig. 5 that we need to go to even higher values of $l$.   Higher values of $R$ also reduce mode-mixing.  Table 3b (table on the right) shows the crosstalk values for  $l=6$ and $R=8cm$. The $OAM_{6,1}$ mode remains impervious to changes in $L$ due to the increased value of the $2\pi$ walk-off length; consequently, there is basically no admixed $OAM_{-6,1}$ content.  The maximum admixed content permissible for $OAM_{\pm5,1}$ mode is $10log_{10}(4|a^{(1)}_{(6,1)(5,1)}|^2) = -21.8  dB$, and $10log_{10}(4|a^{(1)}_{(6,1)(7,1)}|^2) =-23.7 dB$ for  $OAM_{\pm7,1}$ mode. The results in Table 3b satisfy these limits. Additionally, the criterion of maximum permitted crosstalk of $-23 dB$ adopted in Section 6.1.3 is basically satisfied here. If we adopt this criterion  here as well,  then certainly for $R\ge 8 cm$ and $l \ge 6$, we can have   simultaneous propagation of $OAM$ modes with crosstalk $\le -23 dB$. By constructing similar tables followed by analyses, one can  further gain insight into  the impact of the presence of a fiber bend that may exist as fiber coiled up between the input and  output of an OAM mode-multiplexed transmission system. 
\section{Further Remarks}
%
%

\subsection{Consideration of Bend Effect on Polarization and Vector Modes}
In the above derivation, we considered an input $OAM_{l,m}$ mode, but this mode also has a  polarization, left circular $\epsilon_+$ or right circular $\epsilon_-$ assumed here (in general, you can have an arbitrary linear combination). The presence of a bend also changes polarization \cite{smith} with the result that at the output, the polarization, starting initially as a pure $\epsilon_\pm$ becomes a linear combination of $\epsilon_+$ and $\epsilon_-$. This linear combination multiplies the  right-hand-side (RHS) of Eq. 41 resulting in  a linear combination of the spatial-spin product states, $\phi_{l,m}\vec{\epsilon}_+, \phi_{-l,m}\vec{\epsilon}_-, \phi_{l,m}\vec{\epsilon_-}, \phi_{-l,m}\vec{\epsilon_+}$,  that correspond to the vector quartet, $HE_{l+1,m}, HE_{-l-1,m},EH_{l-1,m},$ and $EH_{-l+1,m}$ ($l>1$ ), respectively within the fiber, and similar quartets of spatial-spin product states involving the parameters, $l' (\ne l),m'$ of the admixed states; the latter quartets similarly translate  into their corresponding vector mode quartets. 
\\\\
It is also known from the vector wave equation that spin (S) and orbital angular momentum ($l$) couple within the fiber \cite{snyder,doog,volyar,alex}. As $l$ increases, the spin-orbit $(l-S)$ coupling increases.   At sufficiently  large values of $l$,  the $l-S$ coupling may  be strong enough to lock $l$ and $S$ into each other \cite{rama}; in other words, the spin S will flip sign (change from +1 to -1 and vice versa) only if $l$ changes sign. Consequently, if the bend effect is weak and cannot convert an input $OAM_{l,m}$ mode into  the $OAM_{-l,m}$ mode (i.e., the $2\pi$ walk-off length is large), then because of this locking of $l$ and $S$, the spin or polarization of the input mode would also not change during propagation through the bend.  In terms of vector modes, when the  input polarization is $\epsilon_+$, the spatial $OAM_{l,m}$ mode couples to $HE_{l+1,m}$, and for large values of $l$,   we then expect Eq. 29 to be representative of the difficulty of conversion of $HE_{l+1,m}$ into its negative counterpart, $HE_{-l-1,m}$; similarly, when the input polarization is $\epsilon_-$, the vector mode $EH_{l-1,m}$ is excited and Eq. 29 then, in similar vein,  would describe the difficulty of conversion of $EH_{l-1,m}$ into $EH_{-l+1,m}$.
\subsection{Comparison with Previous Theoretical Work}
The formulas for $2\pi$ walk-off lengths for vector modes like Eq. 27 have been cited earlier \cite{yue,wang}, although without analytic expressions  because  a finite-element solver was employed in the study of the properties of a fiber. The generic formula cited therein pertains to the transformation between the even and odd HE modes and similarly between the even and odd EH modes in the event of a fiber bend \cite{wang} or ellipticity \cite{wang, yue}. From the $l-S$ coupling discussion above,  the transformation between the $HE_{l+1,m}$ and $HE_{-l-1,m}$ vector modes is essentially a transformation between $OAM_{l,m}$ and $OAM_{-l,m}$ modes in the scalar mode theory, at least for large values of $l$. Thus, the $\beta^{'\pm}_{l,m}$ in Eq. 27 may be considered corresponding to  two linear combination of $HE_{l+1,m}$ and $HE_{-l-1,m}$ with their respective coefficients in accordance with Eq. 23. These two linear combinations are the even and odd $HE_{l+1,m}$ modes considered in \cite{wang,yue}. The same argument applies to the $EH_{l-1,m}$ and the $EH_{-l+1,m}$ vector modes.  The generic formula in \cite{wang,yue} is identical with ours (Eq. 27) if we replace each of the propagation constants in Eq. 27 by the corresponding effective refractive index times $k$. 
\\\\
Reference \cite{wang} provides detailed plots of the $2\pi$ walk-off lengths for various vector modes as a function of the bend radius. Although the calculations are performed for a graded-index fiber, we do see the expected rise in the value of the $2\pi$ walk-off length with  increasing  bend radius, the rise also being  sharper for larger $l$ value vector modes.  However, the curves flatten out quickly with increasing bend radius $R$, in the vicinity of $R=2 cm$. This is contrary to expectations because as $R$ becomes large, approaching zero curvature (a straight fiber case), we expect the $2\pi$ walk-off length to approach infinity. This discrepancy is perhaps attributable to the inability of  the software of the finite-element solver (an approximate method) to deal  with small numbers requiring high precision; small numbers occur as effective refractive index differences between the even and odd vector modes in the calculation of  $2\pi$ walk-off length (see Fig. 10a of \cite{wang}).
\\\\
The perturbation theory developed here gives an $R^{2l}$ dependence of the $2\pi$ walk-off length on $R$ consistent with a rising value, which is sharper for larger $l$.  In \cite{wang}, we, however,  see a  case of a $2\pi$ walk-off length being larger for a lower $l$ value (when it should be smaller) and another, where it is larger for a bend radius of $1cm$ compared to one of 
$2 cm$. These discrepancies are perhaps due again to the numerical inaccuracies of the finite-element solver employed. 
%

%

\section{Summary and Conclusion}
We have presented a detailed scalar  perturbation treatment  for the mixing of the OAM modes due to a fiber bend  as a function of bend radius and topological charge $l$. To our knowledge, this is the first such effort.  We employ a well-established  equivalent refractive index model that enables treatment of a bent fiber as a straight fiber. The perturbation analysis  leads to an important  selection rule, namely, that only OAM states differing in topological charge by $\pm1$ can mix in first order of perturbation.  As a result, we are able to gain insight into the mechanism for  mixing of all the modes  including the breaking of the degeneracy between the $+l$ and the $-l$ OAM modes; modes differing in topological charge by $p$ can only be connected in perturbation order $|p|$. The selection rule further yields simplified forms of analytic expressions for mode-mixing in all orders of perturbation. The perturbation parameter is defined to be the ratio of the fiber  core radius to the fiber bend radius, which then provides the dependence of the mixing coefficients on the fiber bend radius; larger the bend radius, lower the value of the mixing coefficient, and vice versa. Analytic expressions for the mixing of the degenerate modes obtain in terms of the $2\pi$ walk-off length, i.e., the distance over which an OAM mode transforms into its negative $l$ counterpart, and back into itself.  Larger the value of $l$, larger the $2\pi$ walk-off length due to the increasing difficulty of a given fiber bend  to provide the required $2l$ transfer of OAM. Furthermore, as the fiber bend radius approaches infinity to become a straight fiber, the $2\pi$ walk-off lengths approach infinity, as one would expect.   Crosstalk (in dB) is defined from the derived analytic expressions.
\\\\
Finally, the derived analytic expressions and their features  are numerically simulated and illustrated with application to a few mode fiber as well as a conventional step-index multimode fiber.   Crosstalk engendered by a fiber coil/spool (of radius $R$ and length $L$) between the input and output of  an OAM mode transmission is calculated and discussed, with implications for a  mode-multiplexed system.   Previous theoretical work pertaining to a graded-index fiber and based on a finite-element solver is also compared with our analytic results. The scalar perturbation theory presented here is general enough to be applicable to other perturbations like ellipticity as well as other fibers which follow a step-like refractive index profile as, for example, in the ring fiber [3].
\\\\
\textbf{Acknowledgement}
\\
  The author is grateful to the referees for their helpful suggestions in improving the paper presentation. He also thanks Ken Ritter   and Tom Salter for their comments and Rusko Ruskov for a technical clarification.

\vspace{0.5cm}
{\large\bf   Appendix}
%
%
%
%
%
\appendix
\numberwithin{equation}{section}
\section {Derivation of the Mixing of the $OAM_{l,m}$ and $OAM_{-l,m}$ Modes due to the Bend}
Because the states $OAM_{l,m}$ and $OAM_{-l,m}$ are degenerate, we need to find the linear combinations, which are orthonormal and appropriate to the perturbation.    Due to the Hermitian nature of $H$ and $\delta H$, these linear combinations are obtained via a unitary transformation in the $(l,-l)$ subspace spanned by the two degenerate modes. Consequently,  we write
\begin{equation}
O^+_{l,m}=U_{11}O_{l,m}+U_{12}O_{-l,m},
\end{equation}
\begin{equation}
O^-_{l,m}=U_{21}O_{l,m}+U_{22}O_{-l,m},
\end{equation}
where $U_{i,j}$ are the elements of the $2 x 2$ unitary matrix $U$.   Let $\beta^{\pm'}_{l,m}$ denote  the eigenvalues  associated with these two linear combinations.
\\\\
We now replace $O_{ l,m}$ in Eq. 6 with $O^{+}_{l,m}$, and  $O_{- l,m}$ in Eq. 19 with  $O^{-}_{l,m}$, and rewrite the two perturbation  series   in a standard procedure \cite{landau,mw,soliverez}  as
\begin{equation}
O'^{\pm}_{l,m}=O^{\pm}_{l,m}+\sum_{n \ne \pm l}\sum_{k}a^{\pm (1)}_{(l,m)(n,k)}O_{n,k} +\sum_{n \ne \pm l}\sum_{k}a^{\pm(2)}_{(l,m)(n,k)}O_{n,k} +...
\end{equation}
The two individual series are labeled by $+$ and $-$ signs,   although $n \ne \pm l$ applies to both the series. The coefficients, $a^{\pm(i)}_{(l,m)(n,k)}$ are defined in the same way as $a_{(l,m)(n,k)}^{(i)}$ (see Eqs. 8 and 9), except that the $\delta H$ matrix elements, which involve the $O_{l,m}$ amplitude in Eq. 12 are replaced with the corresponding matrix elements involving the $O^{\pm}_{l,m}$ amplitudes instead.   Considering the $O^+_{l,m}$ amplitude first, we define 
\begin{equation}
\delta H^+_{(n,k)(l,m)}=<O_{n,k}|\delta H|O^+_{l,m}>,
\end{equation}
 which becomes, upon substitution of Eq. A.1,
\begin{equation}
\delta H^+_{(n,k)(l,m)}=U_{11}\delta H_{(n,k)(l,m)}\delta_{n, l\pm1}+U_{12}\delta H_{(n,k)(-l,m)}\delta_{n,-l\pm 1}.
\end{equation}
Similarly, 
\begin{equation}
\delta H^-_{(n,k)(l,m)}=<O_{n,k}|\delta H|O^-_{l,m}>=U_{21}\delta H_{(n,k)(l,m)}\delta_{n, l\pm1}+U_{22}\delta H_{(n,k)(-l,m)}\delta_{n,-l\pm 1}
\end{equation}
obtains after substituting Eq. A.2.   Replacement of 
 $\delta H_{(n,k)(l,m)}$ with Eqs. A.5 and A.6 in the perturbation series, Eq. A.3, transforms these series as 
\begin{equation}
O'^{\pm}_{l,m}=O^{\pm}_{l,m}+\sum_{n \ne \pm l}\sum_{k} a^{(1)}_{(l,m)(n,k)}O^{\pm}_{n,k}+\sum_{n \ne \pm  l}\sum_{k} a^{(2)}_{(l,m)(n,k)}O^{\pm}_{n,k}+.....,
\end{equation}
where
\begin{equation}
O^+_{n,k}=U_{11}O_{n,k}+U_{12}O_{-n,k},
\end{equation}
\begin{equation}
O^-_{n,k}=U_{21}O_{n,k}+U_{22}O_{-n,k}.
\end{equation}
 Starting from Eq. A.3, we have thus obtained a new form of the series, expressed entirely in terms of linear combinations of the degenerate states. Furthermore, these linear combinations, regardless of the topological charge, are described uniquely by the elements of a single unitary matrix $U$ (to be determined later). The mixing coefficients of $O^{\pm}_{n,k}$ are the same as those in the original perturbation series for $O'_{\pm l,m}$.
\\\\
Consider now
\begin{equation}
(H+\lambda'\delta H)O^{+'}_{l,m}=\beta^{+'2}_{l,m}O^{+'}_{l,m}.
\end{equation}
  We write $\beta^{+'2}_{l,m}=\beta^2_{l,m}+\delta \beta^2_{l,m}$, where $\delta \beta^2_{l,m}$ is an unknown to be determined along with matrix $U$. Making this substitution in Eq. A.10, we obtain
\begin{equation}
(\delta\beta^{2}_{l,m}-\lambda' \delta H)O^{+'}_{l,m}=(H-\beta_{l,m}^2)O^{+'}_{l,m}.
\end{equation}
Taking the scalar product on the left with $O_{l',m} (l'=\pm l)$, we get 
\begin{equation} 
<O_{l',m}|\delta\beta^{2}_{l,m}-\lambda' \delta H|O^{+'}_{l,m}>=<O_{l',m}|H-\beta_{l,m}^2|O^{+'}_{l,m}>=<(H-\beta_{l,m}^2)O_{l',m}|O^{+'}_{l,m}>.
\end{equation}
The last step follows from the Hermiticity of  $H-\beta_{l,m}^2$. Because $(H-\beta_{l,m}^2)O_{l',m}=0$, it follows then
\begin{equation}
<O_{l',m}|\delta\beta^{2}_{l,m}-\lambda' \delta H|O^{+'}_{l,m}>=0.
\end{equation}
 Now the series for $O^{+'}_{l,m}$ (Eq. A.7) can be written as
\begin{equation}
 O^{+'}_{l,m} = O^+_{l,m}+\sum_{p=1}^{\infty}\sum_{n \ne \pm l}\sum_{k} a^{(p)}_{(l,m)(n,k)}O^{+}_{n,k}.
\end{equation}
Inserting Eqs. A.14,  A.1, and A.8 into Eq. A.13, we obtain
\begin{equation}
\begin{split}
&<O_{l',m}|\lambda'\delta H-\delta \beta^2_{l,m}|(U_{11}O_{l,m}+U_{12}O_{-l,m})> \\
&+\sum_{p=1}^{\infty}\sum_{n \ne \pm l}\sum_{k}a^{(p)}_{(l,m)(n,k)}<O_{l',m}|\lambda' \delta H -\delta \beta^2_{l,m}|U_{11}O_{n,k}+U_{12}O_{-n,k})>=0.
\end{split}
\end{equation}
Setting $l'=\pm l$ leads to two linear equations in $U_{11}$ and $U_{12}$:
\\\\
1) \underline{$l'=l$}
\begin{equation}
\lambda'\sum_{p=1}^{\infty}\sum_{n \ne \pm l}\sum_{k}a^{(p)}_{(l,m)(n,k)}(U_{11} \delta H_{(l,m)(n,k)} +U_{12} \delta H_{(l,m)(-n,k)})-U_{11}\delta \beta^2_{l,m}=0.
\end{equation}
\\\\
2) \underline{$l'=-l$}
\begin{equation}
\lambda'\sum_{p=1}^{\infty}\sum_{n \ne \pm l}\sum_{k}a^{(p)}_{(-l,m)(n,k)}(U_{11} \delta H_{(-l,m)(n,k)} +U_{12} \delta H_{(-l,m)(-n,k)})-U_{12}\delta \beta^2_{l,m}=0.
\end{equation}
The above homogeneous equations have a non-trivial solutions if and only if the determinant of the coefficients of $U_{11}$ and $U_{12}$ is zero. Invoking the relationships among the $\delta H$ matrix elements and the fact that the mixing coefficients for $l$ and $-l$ cases are identical in values (see Section 4.2), it is easy to see that the  corresponding matrix  is of the form, $M-\delta \beta^2_{l,m}I$, where matrix $M$ is  symmetric (with diagonal elements equal) and $I$ is the $2 x 2$  identity matrix. For degeneracy to be broken between the $OAM_{l,m}$ and $OAM_{-l,m}$ modes, the off diagonal elements of the $M$ matrix must be non-zero. From an examination of these off diagonal terms (see, e.g., the coefficient of $U_{12}$ in Eq. A.16), and recalling the form of the expressions for $a^{(p)}_{(l,m)(n,k)}$ for $p=1,2,3$ (Eqs. 14-16), and higher, we find that this coefficient is nonzero only when $p=2l-1$. Truncating the infinite series at $p=2l-1$ in Eqs. A.16 and A.17, we  then write the symmetric matrix $M$ as 
\begin{equation}
M=
\begin{bmatrix}
\gamma^{(p)}&\kappa^{(p)}\\
\kappa^{(p)}&\gamma^{(p)}
\end{bmatrix}.
\end{equation}
  The eigenvalue equation is  $(M-\delta \beta^2_{l,m}I)\psi= 0$, where $ \delta\beta^2_{l,m}$ is the eigenvalue of matrix $M$, and $\psi$ (a  $2x1$ column vector) the corresponding eigenvector.  Setting the determinant of ($M-\delta \beta^2_{l,m}I$) to zero then yields the two eigenvalues:
\begin{equation}
\delta \beta^{\pm2}_{l,m}=\gamma^{(p)}\pm\kappa^{(p)}.
\end{equation}
 That is,
\begin{equation}
\beta^{'\pm 2}_{l,m}=\beta^2_{l,m}+\gamma^{(p)}\pm\kappa^{(p)}.
\end{equation}
The corresponding eigenvectors are $\frac{1}{\sqrt{2}}\begin{bmatrix}
1\\
1
\end{bmatrix}$ and  $\frac{1}{\sqrt{2}}\begin{bmatrix}
1\\
-1
\end{bmatrix}$, which are identified with the two column vectors of the $U$ matrix.   That is,  $U_{11}=\frac{1}{\sqrt{2}}$, $U_{12}=\frac{1}{\sqrt{2}}$, $U_{21}=\frac{1}{\sqrt{2}}$, and $U_{22}=-\frac{1}{\sqrt{2}}$. The unitary matrix $U$ is independent of $l$. Inserting   the numerical values of the appropriate $U_{ij}$ elements  in Eqs. A.1 and A.2, we obtain  $O^+_{l,m}=\frac{1}{\sqrt{2}}(O_{l,m}+O_{-l,m})$ and $O^-_{l,m}=\frac{1}{\sqrt{2}}(O_{l,m}-O_{-l,m})$.    Similar insertions in Eqs. A.8 and A.9 yield $O^+_{n,k}=\frac{1}{\sqrt{2}}(O_{n,k}+O_{-n,k})$ and $O^-_{n,k}=\frac{1}{\sqrt{2}}(O_{n,k}-O_{-n,k})$ for $n \ne  l$. As a result, we now know the entire perturbation series, Eq. A.7.   We, however, need to compute $\gamma^{(p)}$ and $\kappa^{(p)}$ also to determine $\beta^{'\pm 2}_{l,m}$. 

 Writing out the coefficient of $U_{11}$ in Eq. A.16, where the series is terminated at $p=2l-1$, and recalling the general form of the expression for expressions for $a^{(p)}_{(l,m)(n,k)}$  (Eqs. 14-16), we immediately see that $\gamma^{(p)}$ (in lowest order of $\lambda'$) is given by
\begin{equation}
\gamma^{(p)}=\lambda^{'2}\sum_{k}\frac{(\delta H_{(l,m)(l-1,k)})(\delta H_{(l-1,k)(l,m)})}{\beta^2_{l,m}-\beta^2_{l-1,k}}
+\lambda^{'2}\sum_{k}\frac{(\delta H_{(l,m)(l+1,k)})(\delta H_{(l+1,k)(l,m)})}{\beta^2_{l,m}-\beta^2_{l+1,k}}.
\end{equation}
This expression is identical to the expression, Eq. 18, for the nondegenerate case, which is not surprising, since $\gamma^{(p)}$ is the diagonal term of the $2x2$ matrix $M$ in the $(l,-l)$ subspace. It is the nonzero nature of the off diagonal term, $\kappa^{(p)}$, which breaks the degeneracy as described above.
\\\\
Similar examination of the coefficient of $U_{12}$ in the terminated series of Eq. A.16 yields for $l=1$  (for which $p=2l-1 =1$)
\begin{equation}
\kappa^{(1)}=\lambda^{'2}\sum_{k}\frac{(\delta H_{(1,m)(0,k)})(\delta H_{(0,k)(-1,m)})}{\beta^2_{1,m}-\beta^2_{0,k}}.
\end{equation}
For $l=2$ (for which $p=3)$
\begin{equation}
\kappa^{(3)}=\lambda^{'4}\sum_{k,i,j}\frac{(\delta H_{(2,m)(1,k)})(\delta H_{(1,k)(0,i)})(\delta H_{(0,i)(-1,j)})(\delta H_{(-1,j)(-2,m)})}{(\beta^2_{2,m}-\beta^2_{1,k})(\beta^2_{2,m}-\beta^2_{0,i})(\beta^2_{2,m}-\beta^2_{1,j})}.
\end{equation}
%

We are primarily interested in $\kappa^{(p)}$ because $\Delta\beta^{'2}_{l,m}=\beta^{'+2}_{l,m}-\beta^{'-2}_{l,m}=2\kappa^{(p)}$ (see Eq. A.20); $p=2l-1$. From the expressions for $\kappa^{(1)}$ and $\kappa^{(3)}$, and a further analysis,  a generalization emerges, and we obtain Eq. 22   for arbitrary $l$.

\end{document}